\documentclass[prb,twocolumn,showpacs,preprintnumbers,
amsmath,amssymb]{revtex4}

\usepackage{graphicx} 
\usepackage{dcolumn} 
\usepackage{bm}
\usepackage{txfonts}
\usepackage{mathrsfs}
\usepackage{feynmf}
\usepackage{comment}

\usepackage[]{psfrag}

\psfrag{tyldu}{$\tilde{u}$}
\psfrag{deltal}{$\delta$}
\psfrag{logdelt}{$\log |a_2-a_2^{(0)}|$}
\psfrag{logphi}{$\log (\phi_0)$}
\psfrag{dgdc}{$\frac{\delta_G-\delta_c}{\delta_c}$}
\psfrag{Temp}{$T$}
\psfrag{etaom}{$\eta_\omega$}
\psfrag{aa22}{$a_2$}
\psfrag{aa44}{$a_4$}
\psfrag{rown1}{$a_4^2=3a_2a_6$}
\psfrag{rown2}{$a_4^2=4a_2a_6$}
\psfrag{logTTT}{$\log(T^{tri}-T)$}
\psfrag{rhobar}{\hspace{0.5cm}$\bar{\rho}$}

\begin{document}


\title{Renormalized $\phi^6$ model for quantum phase transitions in systems of itinerant fermions}

\author{P. Jakubczyk}
\email{p.jakubczyk@fkf.mpg.de}
\affiliation{Max-Planck-Institute for Solid State Research,
Heisenbergstr.~1, D-70569 Stuttgart, Germany } 
\affiliation{Institute for Theoretical Physics, Warsaw University, 
 Ho\.za 69, 00-681 Warsaw, Poland}

\date{\today}

\begin{abstract}
We study the impact of quantum and thermal fluctuations on properties of
quantum phase transitions occurring in systems of itinerant fermions with main focus on
the order of these transitions. Our approach is based on a set of flow
equations derived within the functional renormalization group framework, in which
the order parameter is retained as the only degree of freedom, and where the
effective potential is parametrized with a $\phi^6$ form allowing for both first and
second order scenarios. We find a tendency to turn the first order transitions within the bare model into second order transitions upon accounting for the order parameter fluctuations. We compute the first and second order phase boundary lines $T_c$ as a function of a non-thermal control parameter $a_2$ in the vicinity of a quantum phase transition. We analyze crossovers of the shift exponent $\psi$ governing the shape of the $T_c$ line when the system is tuned close to a quantum tricritical scenario, where a second order phase transition line terminates at a quantum tricritical point.
\end{abstract}

\maketitle

\section{Introduction}
Quantum phase transitions in systems of itinerant fermions continue to attract considerable interest.\cite{sachdev_book, loehneysen_review06, belitz_review05, abanov03, gegenwart08} On one hand this attention is due to the relevance of quantum critical points for understanding collective phenomena like high-$T_c$ superconductivity and other non-conventional properties of experimentally investigated compounds.\cite{stewart01} On the other, it is related to peculiarities that so far prohibited the emergence of a fully convincing theoretical description of most systems exhibiting quantum criticality.\cite{loehneysen_review06, belitz_review05}

The standard approach in describing quantum critical phenomena in systems of itinerant fermions - the Hertz-Millis theory \cite{hertz76, millis93} - relies on an order parameter field, which is introduced by a Stratonovich-Hubbard transformation. In a subsequent step  the fermionic degrees of freedom are integrated out. The resulting representation of the partition function as a path integral over a bosonic field is usually useful only provided the action may be expanded in the order parameter, and the relevant vertex functions can be evaluated for the values of momentum and frequency that correspond to the anticipated instability. The validity of such an expansion was questioned for both zero and finite momentum instabilities in the magnetic channel. In case of ferromagnetic transitions it was argued that the correct effective action involves additional terms \cite{belitz97, chubukov03, chubukov04} leading to a first order transition, \cite{belitz99} while for the case of antiferromagnetic transition in $d=2$ integrating out gapless fermionic modes can yield singular vertex functions all of which were argued to be marginal. \cite{abanov_chubukov04} Despite these theoretical insufficiencies, Hertz-Millis theory is successful in explaining a number of nonconventional properties of many systems exhibiting quantum criticality. \cite{stewart01}  In recent years this theory was further extended to account for a number of systems not described by the original approach. These include metamagnetic quantum critical points,\cite{millis_schofield02} field-tuned quantum critical points where a term describing precession of the order parameter has to be retained in the action,\cite{fischer_rosch05} and phase transitions induced by a nonequilibrium drive. \cite{mitra_takei06} 

An interesting issue concerns the actual order of specific quantum phase transitions. Hertz-Millis theory by construction allows for continuous transitions only. However, microscopic models, which usually rely on fermionic degrees of freedom and meanfield-like treatment, often predict a first order scenario in case of both magnetic and charge instabilities occurring at $Q=0$, and for specific cases corresponding to $Q\neq 0$. \cite{altshuler95} 
Theoretical studies of the interplay between first and second order scenarios are further motivated by experimental results, where the order of the quantum phase transitions in specific compounds is altered by varying external magnetic field $h$.\cite{pfleiderer01, uhlarz04} Upon increasing $h$, the tricritical point separating the first order transition at $T<T^{tri}$ from the second order at $T>T^{tri}$, is shifted towards lower values of $T$ and finally vanishes at $T=0$. Therefore, the order of the transition at small $T$ can be  controlled by varying $h$.  Of particular interest is the case $T^{tri}=0$, where the transition is second order for all $T>0$, but scaling behavior of physical quantities at low temperatures is in variance with that predicted by Hertz-Millis theory. \cite{belitz05, green05, misawa08} An interesting scenario of quantum criticality occurs also in systems exhibiting a metamagnetic first order transition terminating with a critical end point.\cite{millis_schofield02, grigera_et_al_01} By tuning pressure the critical end point is suppressed to $T=0$ resulting in a distinct type of a quantum critical point.

In this work we investigate the possibility of altering the order of quantum phase transitions by order parameter fluctuations. Subsequently, we study shapes of phase boundaries, focusing mainly on the cases where such change occurs. 
We rely on the conventional bosonic Hertz action adapted to phases with broken symmetry and retaining also a $\phi^6$ term in the effective potential. We analyze two- and three- dimensional systems where discrete symmetry-breaking occurs at $Q=0$ (the dynamical exponent $z=3$). Our results should apply whenever an action of this type can be constructed.
Possible examples are magnetic transitions with Ising-like symmetry and the so-called Pomeranchuk instabilities \cite{yamase00, halboth00} where the discrete point-group lattice symmetry of the Fermi surface is broken. For these transitions mean-field studies typically predict a first order scenario at $T=0$ but second order for slightly higher temperatures (for $T>T^{tri}$).\cite{kee03, khavkine04, yamase05} We argue that the tricritical temperature is lowered by including the order-parameter fluctuations. Under appropriate conditions the tricritical point may be suppressed to $T=0$, in which case a quantum critical or tricritical point is realized. 

Our study is based on the one-particle irreducible version of the functional renormalization group (RG).\cite{wetterich93, berges_review02, delamotte07, metzner_review, pawlowski_review_07, gies_notes_06} This framework was applied extensively in the context of classical critical phenomena, where it yields a unified description of $O(N)$-symmetric scalar models including systems in $d=2$.\cite{berges_review02} It also provides a suitable framework to treat quantum criticality.\cite{wetterich08, jakubczyk08}  We apply the derivative expansion to quadratic order, and the effective potential is parametrized in a form allowing for the occurrence of both - first and second order transition. An advantage of the functional RG is that it allows control over effects occurring at different energy scales, and treatment of quantities that need not be universal.

The outline of this work is as follows: In Sec. II we introduce the bare action to be applied in the RG procedure and provide a phase diagram of the unrenormalized theory.  In Sec. III we discuss the applied truncation of the functional RG flow equation and derive equations governing the evolution of quantities parametrizing the effective potential and the inverse propagator upon reducing the cutoff scale. Sec. IV is devoted to results obtained for the renormalized phase diagrams in the case $T=0$. In Sec. V we consider $T>0$. We discuss the computed phase diagrams for cases where the phase  transition is second order for $T>0$ and terminates at a quantum critical or tricritical point at $T=0$. We also analyze the case where a tricritical point situated at $T>0$ occurs. For $d=3$ and a choice of system parameters corresponding to proximity to the quantum tricritical scenario we compute the crossovers of the shift exponent $\psi$, characterizing the shape of the critical line, from the Hertz-Millis value $\psi^{HM}=3/4$ to $\psi^{tri}=3/8$. This latter value of the shift exponent is specific to quantum tricritical behavior in $d=z=3$. The obtained value of $\psi^{tri}$ is justified by invoking a scaling argument in Sec. VI, where we also classify and discuss scaling regimes emergent from our analysis. Finally, Sec. VII contains a summary of the work. 

\section{Bare action}
In this study we rely on the conventional description of quantum critical points in systems of itinerant fermions in terms of the bosonic Hertz action.\cite{hertz76} This is derived by applying a Stratonovich-Hubbard transformation to the path-integral representation of the partition function of a fermionic system.\cite{negele_orland} This way the order-parameter field is introduced and the original fermionic degrees of freedom are decoupled allowing one to integrate them out. Subsequently, one expands the resulting action in powers of the order-parameter field keeping only the most relevant dependence of the propagator on frequency and momentum, and neglecting this dependence in case of higher-order vertices. In the case of instabilities occurring at wavevector $Q=0$, which is considered here, this leads to the following action
 \begin{eqnarray}
 S[\phi] = 
 \frac{T}{2} \sum_{\omega_n} \int\frac{d^dp}{(2\pi)^d} \,
 \phi_p \left( \frac{|\omega_{n}|}{|\mathbf{p}|}
 + \mathbf{p}^{2} \right) \phi_{-p} + U[\phi] \; .
 \label{eq:lagrangian}
\end{eqnarray}
Here $\phi$ is the scalar order parameter field and $\phi_p$ with
$p = (\mathbf{p},\omega_n)$ its momentum representation;
$\omega_n = 2\pi n T$ with integer $n$ denotes the (bosonic) 
Matsubara frequencies. 
Momentum and energy units are chosen such that the prefactors 
in front of $\frac{|\omega_{n}|}{|\mathbf{p}|}$ and 
$\mathbf{p}^{2}$ are equal to unity. 
The action is regularized in the ultraviolet by
restricting momenta to $|\mathbf{p}| \leq \Lambda_0$. The term $\frac{|\omega_{n}|}{|\mathbf{p}|}$ effectively accounts for overdamping of the order-parameter fluctuations by fermionic excitations across the Fermi surface. The expression Eq.~(\ref{eq:lagrangian}) is valid for $|\mathbf{p}|$, and $\frac{|\omega_{n}|}{|\mathbf{p}|}$ sufficiently small, which is the limit relevant for the physical situation considered here.\cite{millis93}

The potential $U[\phi]$ is usually parametrized by a quartic form with a single minimum at $\phi=0$ or with two minima at  $\phi=\pm\phi_0\neq 0$.  As we wish to allow for both, first and second order transitions, we expand $U[\phi]$ to sixth order in $\phi$, imposing 
\begin{equation}
 U[\phi]=\int_0^{1/T}  d\tau \int \! d^d x
 \left[a_6\phi^6+a_4\phi^4+a_2\phi^2\right].
\label{eq:ef_potential}
\end{equation}
We consider $a_6>0$ to stabilize the system at large $|\phi|$. 

In Eq.~(\ref{eq:ef_potential}) we assumed the existence of an expansion of $U[\phi]$, which correctly captures the properties of $U[\phi]$ also around the non-zero minima. One should be aware, that in many physical situations such expansion necessarily has to involve also terms of higher order in $\phi$, or simply cannot be constructed. The latter, more severe case occurs whenever the radius of convergence around $\phi=0$ is smaller than the distance between $\phi=0$ and the minimum at $\phi=\phi_0\neq 0$. Our belief is that except for the vicinity of the tricritical point, such complication is rather generic. It arises for example in case of models describing symmetry-breaking Fermi surface deformations in $d=2$ (see e.g. Ref.~\onlinecite{yamase05}).  While the model defined by Eq.~(\ref{eq:ef_potential}) is one of the simplest correctly accounting for universal critical and tricritical properties, its applicability in the context of specific microscopic models has to be verified for each individual case in the regime of parameters where a first-order transition occurs. Also note, that when $U$ is expanded around the non-zero minimum $\phi_0$, terms involving odd powers of field are generated. The coefficients of these terms can be expressed as functions of the coupling $a_2$, $a_4$, $a_6$. There arise just three independent couplings because of the assumed inversion symmetry $U[-\phi]=U[\phi]$. This model does not apply to physical situations where the symmetry of $U$ is explicitly broken by including a field coupling linearly to $\phi$. For example the present work does not apply to the metamagnetic transitions. \cite{millis_schofield02, grigera_et_al_01}   

The bare effective potential is now analyzed as a function of the parameters $a_2$, $a_4$, $a_6$. It exhibits a single minimum for $a_2>0$, $a_4^2<3a_2a_6$, or $a_2>0$, $a_4>0$. For $a_2<0$ there are two minima located at 
\begin{equation}
\pm\phi_0=\pm\sqrt{\frac{1}{3a_6}\left(-a_4+\sqrt{a_4^2-3a_2a_6}\right)}\;.
\label{phi_0}
\end{equation}
 For $a_4<0$, $a_4^2>3a_2a_6$, one encounters three minima at $\phi=0$ and $\phi=\pm\phi_0$. If in addition $a_4^2>4a_2a_6$ holds, $U[\phi_0]<U[0]$ and a configuration corresponding to $\pm\phi_0$ is stable, otherwise $U[\phi_0]>U[0]$ and $\phi=0$ is the stable state. Along the line $a_2=0$, $a_4>0$ a second-order phase transition occurs, while a first-order transition line is located along the parabola $a_4^2=4a_2a_6$ for $a_4<0$. The tricritical point occurs at $a_2=a_4=0$. This simple analysis of the bare phase diagram is summarized in Fig.1.

\begin{figure}[ht!]
\begin{center}
\includegraphics[width=3.2in]{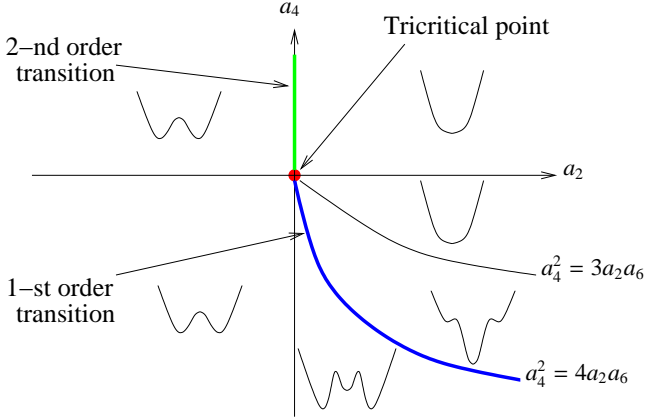}
\caption{(Color online) Schematic phase diagram of the bare $\phi^6$ model. The line $a_2=0$, $a_4>0$ marks the locus of second-order phase transitions, while the curve $a_4^2=4a_2a_6$, $a_4<0$ corresponds to a first order transition. The tricritical point is located at $a_2=a_4=0$. The curves $a_2=0$ and $a_4^2=3a_2a_6$ for $a_4<0$ mark boundaries of the region where metastable states are present.}
\label{fig:eta_vs_lambda}
\end{center}
\end{figure}
In course of the subsequent analysis we investigate the stability of this phase diagram upon including fluctuations, depending on the system dimensionality (for $d=2,3$) and temperature (for $T\geq 0$). 
\section{Flow equations}
 Our analysis is based on approximate flow equations derived by truncating the exact functional RG evolution equation. We use the one-particle irreducible scheme. The flow equation \cite{wetterich93}
\begin{eqnarray}
\frac{d}{d \Lambda}\Gamma^{\Lambda}\left[\phi\right]=
\frac{1}{2}\text{Tr}\frac{\partial_\Lambda R^{\Lambda}}{\Gamma^{(2)}\left[\phi\right]
+ R^{\Lambda}} 
\label{eq:flow_eqn}
\end{eqnarray}
describes the evolution of the effective action $\Gamma^\Lambda[\phi]$, the generating functional for one-particle irreducible vertex functions in the presence of an infrared cutoff at scale $\Lambda$, under reducing the cutoff scale. In Eq.~(\ref{eq:flow_eqn}) $R^\Lambda$ denotes the cutoff function added to the inverse propagator to cut off modes with momentum below the scale $\Lambda$, and
$\Gamma^{(2)}\left[\phi\right] = \delta^{2}\Gamma^{\Lambda}[\phi]/
 \delta \phi^{2}$. In momentum representation the trace sums over momenta and Matsubara frequencies, that is $\text{Tr} = T \sum_{\omega_{n}} 
 \int \frac{d^{d} p}{\left(2\pi\right)^{d}}$. In what follows, we use a regulator term with the optimized Litim cutoff function \cite{litim01} 
\begin{equation}
 R^{\Lambda}(\mathbf{p}) = 
 Z \left( \Lambda^{2}-\mathbf{p}^{2}\right)
 \theta\left(\Lambda^{2}-\mathbf{p}^{2} \right) \; ,
\label{Litim_fun}
\end{equation}
where $Z$ is the wave function renormalization.

The effective action interpolates smoothly between the bare action Eq.~(\ref{eq:lagrangian}) for $\Lambda=\Lambda_0$, and the full effective action for $\Lambda \to 0$. The quantity $\Gamma^\Lambda [\phi]$ can be interpreted as the Gibbs free energy of the rapid modes, i.e. modes with momentum above the cutoff scale $\Lambda$. By decreasing $\Lambda$ fluctuations of lower momentum are included and in the limit $\Lambda\to 0$ the full Gibbs free energy functional is recovered. The general strategy employed in this paper for computing phase diagrams amounts to evaluating $\lim_{\Lambda\to 0}\Gamma^\Lambda [\phi]$ as function of the (bare) parameters describing the system. The equilibrium order parameter is extracted by finding the global minimum of the resulting free energy. Eq.~(\ref{eq:flow_eqn}) describes the flow of an infinite number of couplings and to make progress we shall impose a suitable parametrization in terms of a finite number of variables. Namely, we apply a truncation in which the effective potential $U[\phi]$, that is $\Gamma^\Lambda [\phi]$ evaluated for momentum-independent field,  is taken as a sixth-order polynomial as in Eq.~(\ref{eq:ef_potential}), with flowing coefficients, while the inverse propagator is assumed to maintain the form due to Hertz, as in Eq.~(\ref{eq:lagrangian}), where the coefficients in front of $\frac{|\omega_{n}|}{|\mathbf{p}|}$ and $\mathbf{p}^{2}$ may be renormalized when necessary. Thus, we put 
\begin{equation}
G^{-1}(\textbf{p},\omega_n) = \Gamma^{(2)}\left[\phi=\phi_0\right] = 
Z_{\omega} \frac{|\omega_{n}|}{|\mathbf{p}|} + 
 Z\mathbf{p}^2 + 2a_2 + R^{\Lambda}(\textbf{p}) \; .
\label{Green}
\end{equation}
The quantities $Z$, $Z_\omega$ depend on the scale $\Lambda$ only. Also note, that the allowed forms of the effective potential, as illustrated in Fig.~\ref{fig:eta_vs_lambda}  are typical to $\phi^6$ like truncations. In particular, if no truncation of the functional RG flow equation was performed, the effective potential should obey bounds on the curvature at $\rho=0$,\cite{berges_review02} and become convex for $\Lambda\to 0$. 

An analogous truncation (retaining terms up to quartic order in the effective potential) was applied in Ref.~\onlinecite{jakubczyk08}, where it yields a simple extension of Hertz-Millis theory to phases with broken discrete symmetry, capturing the non-Gaussian fixed point behavior at finite temperatures. 

\subsection{Effective potential flow}
\begin{figure}
\begin{fmffile}{20070802_loops_1}
\begin{eqnarray}
a_2 &:&
\parbox{25mm}{\unitlength=1mm\fmfframe(2,2)(1,1){
\begin{fmfgraph*}(15,20)\fmfpen{thin} 
 \fmfleft{l1}
 \fmfright{r1}
 \fmftop{v1}
 \fmfpolyn{full,tension=0.6}{G}{4}
 \fmf{dbl_wiggly,straight}{l1,G4}
 \fmf{dbl_wiggly,straight}{G1,r1}
 \fmffreeze
\fmf{dbl_wiggly,tension=0.1,right=0.7}{G2,v1}
\fmf{dbl_wiggly,tension=0.1,right=0.7}{v1,G3}
\end{fmfgraph*}
}}
+
\parbox{25mm}{\unitlength=1mm\fmfframe(2,2)(1,1){
\begin{fmfgraph*}(20,18)\fmfpen{thin} 
 \fmfleft{l1}
 \fmfright{r1}
 \fmfpolyn{full,tension=0.3}{G}{3}
 \fmfpolyn{full,tension=0.3}{K}{3}
  \fmf{dbl_wiggly}{l1,G1}
 \fmf{dbl_wiggly,tension=0.2,right=0.8}{G2,K3}
 \fmf{dbl_wiggly,tension=0.2,right=0.8}{K2,G3}
 \fmf{dbl_wiggly}{K1,r1}
 \end{fmfgraph*}
}}\nonumber\\[-1mm]
a_4 &:&
\parbox{25mm}{\unitlength=1mm\fmfframe(2,2)(1,1){
\begin{fmfgraph*}(20,12)
\fmfpen{thin}
\fmfleftn{l}{2}\fmfrightn{r}{2}
\fmfrpolyn{full,tension=0.9}{G}{4}
\fmfpolyn{full,tension=0.9}{K}{4}
\fmf{dbl_wiggly}{l1,G1}\fmf{dbl_wiggly}{l2,G2}
\fmf{dbl_wiggly}{K1,r1}\fmf{dbl_wiggly}{K2,r2}
\fmf{dbl_wiggly,left=.5,tension=.2}{G3,K3}
\fmf{dbl_wiggly,right=.5,tension=.2}{G4,K4}
\end{fmfgraph*}
}}
+
\parbox{25mm}{\unitlength=1mm\fmfframe(2,2)(1,1){
\begin{fmfgraph*}(20,12)
\fmfpen{thin}
\fmfleftn{l}{2}\fmfrightn{r}{2}
\fmfrpolyn{full,tension=1.1}{V}{4}
\fmfpolyn{full,tension=0.5}{W}{3}
\fmfpolyn{full,tension=0.5}{Y}{3}
\fmf{dbl_wiggly}{l1,V1}\fmf{dbl_wiggly}{l2,V2}
\fmf{dbl_wiggly}{W1,r1}\fmf{dbl_wiggly}{Y1,r2}
\fmf{dbl_wiggly,left=.5,tension=.4}{V3,Y2}
\fmf{dbl_wiggly,right=.5,tension=.4}{V4,W3}
\fmf{dbl_wiggly,right=.5,tension=.4}{W2,Y3}
\end{fmfgraph*}
}}
\nonumber\\[+2mm]
&+&
\parbox{25mm}{\unitlength=1mm\fmfframe(2,2)(1,1){
\begin{fmfgraph*}(20,12)
\fmfpen{thin}
\fmfleftn{l}{1}\fmfrightn{r}{3}
\fmfrpolyn{full,tension=0.4}{V}{3}
\fmfpolyn{full,tension=1.3}{W}{5}
\fmf{dbl_wiggly}{l1,V1}
\fmf{dbl_wiggly}{W1,r1}\fmf{dbl_wiggly}{W2,r2}\fmf{dbl_wiggly}{W3,r3}
\fmf{dbl_wiggly,left=.5,tension=.4}{V2,W4}
\fmf{dbl_wiggly,right=.5,tension=.4}{V3,W5}
\end{fmfgraph*}
}}
+
\parbox{25mm}{\unitlength=1mm\fmfframe(2,2)(1,1){
\begin{fmfgraph*}(15,10)\fmfpen{thin} 
 \fmfsurroundn{v}{6}
 \fmfpolyn{full,tension=2.0}{G}{6}
 \fmf{dbl_wiggly,left=0.5}{v1,G1}
 \fmf{dbl_wiggly,right=0.5}{v4,G4}
 \fmf{dbl_wiggly,straight,tension=0.2}{v5,G5}
 \fmf{dbl_wiggly,straight,tension=0.2}{v6,G6}
\fmf{dbl_wiggly,tension=0.1,right=0.7}{G2,v2}
\fmf{dbl_wiggly,tension=0.1,right=0.7}{v2,v3}
\fmf{dbl_wiggly,tension=0.1,right=0.7}{v3,G3}
\end{fmfgraph*}
}}
\nonumber\\[+3mm]
a_6 &:&
\parbox{25mm}{\unitlength=1mm\fmfframe(2,2)(1,1){
\begin{fmfgraph*}(20,12)
\fmfpen{thin}
\fmfleftn{l}{2}\fmfrightn{r}{4}
\fmfrpolyn{full,tension=1.4}{V}{4}
\fmfpolyn{full,tension=2.0}{W}{6}
\fmf{dbl_wiggly}{l1,V1}\fmf{dbl_wiggly}{l2,V2}
\fmf{dbl_wiggly}{W1,r1}\fmf{dbl_wiggly}{W2,r2}\fmf{dbl_wiggly}{W3,r3}\fmf{dbl_wiggly}{W4,r4}
\fmf{dbl_wiggly,left=.5,tension=.4}{V3,W5}
\fmf{dbl_wiggly,right=.5,tension=.4}{V4,W6}
\end{fmfgraph*}
}}
+
\parbox{25mm}{\unitlength=1mm\fmfframe(2,2)(1,1){
\begin{fmfgraph*}(20,15)
\fmfpen{thin}
\fmfleftn{l}{4}\fmfrightn{r}{4}\fmfbottomn{b}{4}
\fmfrpolyn{full,tension=1.1}{V}{4}
\fmfpolyn{full,tension=1.1}{W}{4}
\fmfpolyn{full,tension=0.8}{Y}{4}
\fmf{dbl_wiggly}{l2,V1}\fmf{dbl_wiggly}{l3,V2}
\fmf{dbl_wiggly}{W1,r2}\fmf{dbl_wiggly}{W2,r3}
\fmf{dbl_wiggly}{Y1,b2}\fmf{dbl_wiggly}{Y2,b3}
\fmf{dbl_wiggly,left=0.7,tension=.4}{V3,W3}
\fmf{dbl_wiggly,tension=.4}{V4,Y4}
\fmf{dbl_wiggly,tension=.4}{W4,Y3}
\end{fmfgraph*}
}}
\nonumber\\[-5mm]\nonumber
\end{eqnarray}
\end{fmffile}
\caption{Feynman diagrams representing the contributions to the flow 
 equations for the couplings parametrizing the effective potential.}
\label{Feynm_diag}
\end{figure}
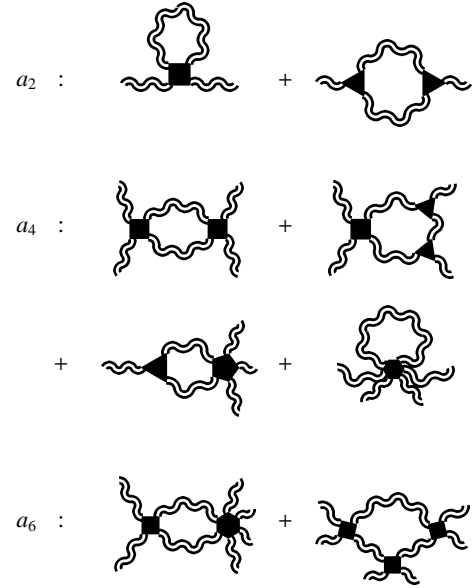

Upon evaluating Eq.~(\ref{eq:flow_eqn}) for a momentum-independent field $\phi$ 
we obtain an exact equation governing the flow of the effective potential $U[\phi]$ (see Ref.~\onlinecite{berges_review02})
\begin{equation}
 \partial_t U[\rho]=2v_d\Lambda^d l_0^d(w)\;,
\label{ef_pot_flow}
\end{equation}
where $\rho=\frac{1}{2}\phi^2$, and 
\begin{equation}
 l_0^d(w)=\frac{1}{4}v_d^{-1}\Lambda^{-d}\text{Tr} \, \frac{\partial_t R^{\Lambda}(\mathbf{p})}
 {Z_{\omega}\frac{|\omega_n|}{|\textbf{p}|} + 
 Z\textbf{p}^2 + R^\Lambda(\mathbf{p}) + Z\Lambda^2w}\; .
\label{l0}
\end{equation}
Here $v_d^{-1}=2^{d+1}\pi^{d/2}\Gamma(d/2)$,  $w=w(\rho)=\frac{1}{Z\Lambda^2}(U'[\rho]+2\rho U''[\rho])$, and $t=\log(\Lambda/\Lambda_0)$. For our parametrization, the partial differential equation Eq.~(\ref{ef_pot_flow}) can be projected onto a set of three ordinary differential equations. Following Ref.~\onlinecite{berges_review02} we neglect $\partial_{\Lambda}Z$ in $\partial_{\Lambda}R^{\Lambda}$. The neglected terms would yield corrections to the flow equations that are linear in $\eta$ and therefore irrelevent except close vicintiy of second order transition in $T>0$. In the latter case these corrections are small as compared to other included terms involving $\eta$ due to the presence of additional factors involving the interaction couplings and mass. We use the condition $U'[\rho_0] = 0$, to write
$0 = \frac{d}{dt} U'[\rho_0] = 
 \partial_t U'[\rho_0] +
 U''[\rho_0] \, \partial_t\rho_0$.
Inserting $\partial_t U'[\rho_0]$ as obtained by differentiating 
Eq.~(\ref{ef_pot_flow}) with respect to $\rho$ at $\rho = \rho_0$,
one obtains the flow equation for $\rho_0$
\begin{equation}
\partial_t \rho_0=2v_d Z^{-1}\Lambda^{d-2}\left(3+2\frac{6a_6\rho_0}{6a_6\rho_0+a_4}\right)l_1^d(w)|_{\rho=\rho_0}\;,
\label{flow_rho_0}
\end{equation}
where the threshold functions $l_i^d(w)$ are defined by
\begin{equation}
 l_1^d(w)=-\frac{\partial}{\partial w}l_0^d(w)\; ,
\end{equation}
\begin{equation}
l_n^d(w)=-\frac{1}{n-1}\frac{\partial}{\partial w}l_{n-1}^d(w)\;,\;\;\; n\geq 2\;.
\end{equation}
From Eq.~(\ref{flow_rho_0}) we conclude that $\rho_0$ decreases under the flow. This follows from $6a_6\rho_0+a_4=\sqrt{a_4^2-3a_2a_6}>0$, and $l_1^d(w)>0$. 
Analogously, by evaluating the second and third derivatives of Eq.~(\ref{ef_pot_flow}) at $\rho_0$ and using $U''[\rho]=48a_6\rho+8a_4$, $U'''[\rho]=48a_6$, which follows from the ansatz, Eq.~(\ref{eq:ef_potential}), we project out the equations governing the flow of $a_4$ and $a_6$: 
\begin{equation}
\partial_t a_4 = 12v_d\Lambda^d\left[\frac{4}{3}l_2^d\frac{(30a_6\rho_0+3a_4)^2}{Z^2\Lambda^4}-5l_1^d\frac{a_6}{Z\Lambda^2}\right]-6\rho_0\partial_t a_6\;,
\label{flow_a4}
\end{equation}
\begin{equation}
\partial_t a_6 = 16v_d\Lambda^d\left[-\frac{8}{3}l_3^d\frac{(30a_6\rho_0+3a_4)^3}{Z^3\Lambda^6}+15l_2^d a_6\frac{30a_6\rho_0+3a_4}{Z^2\Lambda^4}\right],
\label{flow_a6}
\end{equation}
where the threshold functions are evaluated at $w$ corresponding to $\rho_0$, that is $l_n^d=l_n^d(w)|_{\rho=\rho_0}$. The flowing coupling $a_2$ is calculated using Eqs.~(\ref{flow_rho_0}, \ref{flow_a4}, \ref{flow_a6}), and Eq.~(\ref{phi_0}), which yields
\begin{equation}
 a_2=-4\rho_0(a_4+3a_6\rho_0)\;.
\label{a_2}
\end{equation}
The contributions to the flow of the couplings $a_2$, $a_4$, $a_6$ are illustrated in terms of Feynman diagrams in Fig.~\ref{Feynm_diag}.

The threshold functions can be split into classical contributions, that is contributions from $\omega=0$, and the quantum parts. The Matsubara sums are then evaluated analytically, yielding the following expression for the function $l_1^d$:
\begin{equation}
 l_1^d=l_1^{d(cl)}+l_1^{d(q)}\;,
\end{equation}
where 
\begin{equation}
l_1^{d(cl)} = \frac{2}{d}\frac{T}{(1+w)^2}\;, 
\end{equation}
and
\begin{equation}
l_1^{d(q)}=\frac{2T}{\tilde{T}^2}\int_0^1dyy^{\frac{1}{2}(d+2z)-3}\psi_1\left(1+\frac{1+w}{\tilde{T}}y^{\frac{z}{2}-1}\right)\;.
\end{equation}
Here we substituted $y=p^2/\Lambda^2$ and $\tilde{T}=\frac{2\pi TZ_\omega}{Z\Lambda^3}$. The polygamma function $\psi_1$ originated from evaluating Matsubara sum of the type 
\begin{equation}
\sum_{n=1}^{\infty}\frac{1}{(n+C)^2}=\psi_1(C+1)\;.
\end{equation}
By taking derivatives with respect to $w$ and using $\psi_n'(C)=\psi_{n+1}(C)$, we evaluate the higher order threshold functions. For any $l_n^d$ one finds a classical contribution proportional to $T$, and therefore vanishing in the zero temperature limit; and a quantum contribution that survives the limit $T\to 0$, however, at finite $T$, vanishes quickly as the cutoff scale is reduced. 

One technical remark is in place here. In the derivation sketched above it is assumed that $U''[\rho_0]\neq 0$. This condition is not fulfilled whenever the minimum of $U[\rho]$ at $\rho_0\neq 0$ disappears in course of the flow, that is the metastable states are swept away by fluctuations. Indeed, by inspecting Eq.~(\ref{flow_rho_0}) we realize that in such case a singularity in $\partial_t\rho_0$ develops for some finite $t$. While solving the flow equations (see Sec. IV, V) we will simply terminate the flow whenever such singularity is encountered and conclude that the system is in the symmetric phase with no metastable states present, that is we assume that once the metastable states at $\rho_0$ are eliminated by fluctuations, they would not appear again.

At this point it is also worthwhile making a comment concerning some qualitative properties of Eq.~(\ref{flow_a4}). Its right-hand side contains terms proportional to powers of $(30a_6\rho_0+3a_4)$ and a single negative term involving $a_6$. In the language of Feynman diagrams, this term originates from contracting the external legs of the 6-point vertex with a single-scale propagator (the last diagram contributing to the flow of $a_4$ in Fig.~\ref{Feynm_diag}).  Once we choose the mean field (bare) tricritical point coordinates $\rho_0=0$, $a_4=0$ as the initial condition of the evolution equations Eqs.~(\ref{flow_rho_0}, \ref{flow_a4}, \ref{flow_a6}), only this term survives, and the system is driven into a region of the phase diagram with $a_4>0$, where a second-order transition occurs for some value of $a_2$. The same happens if one starts with a negative, though sufficiently close to zero value of the coupling $a_4$. This tendency occurs only in the initial stage of the flow. However, it is unlikely to be inverted in later stages of the evolution, because $\partial_t a_4$ vanishes quickly as $\Lambda\to 0$, thus, significant renormalizations occur mainly at initial stages of the flow - see Sec. IV and V. 
\subsection{Propagator flow}
Within the truncation proposed here, the inverse propagator is parametrized with the flowing $Z$-factors. In what follows, we will in addition neglect the renormalization of $Z_\omega$. As was argued in Ref.~\onlinecite{jakubczyk08}, this is not expected to influence the results for the phase diagram. Renormalization of the factor $Z$ will be considered only at $T>0$, where non-Gaussian critical behavior occurs. The evolution equation for $Z$ is obtained by considering the Laplacian of the inverse propagator
\begin{equation}
 \partial_t Z = \left.
 \frac{1}{2d} \, \Delta_{\textbf{p}}
 \left[\partial_t G^{-1}(\textbf{p},\omega_n=0) \right] 
 \right|_{\textbf{p} = 0} \; ,
\label{Zp}
\end{equation}
and plugging in the flow equation for $G^{-1}(p)$:
\begin{equation}
 \partial_t G^{-1}(p) = 
 (120a_6\phi_0^3+24a_4\phi_0)^2 \text{Tr} \left[ \partial_t R^{\Lambda}(\mathbf{q}) 
 G^2(q) \, G(q+p) \right].
\label{Gamma2flow}
\end{equation}
The latter is obtained by evaluating the second functional derivative of Eq.~(\ref{eq:flow_eqn}). In our calculations we use the order parameter anomalous dimension, related to $Z$ by 
\begin{equation}
 \eta = - \frac{d\log Z}{d\log\Lambda} \; .
\label{eta}
\end{equation}
By carrying out the Matsubara sums one arrives at the following expression for $\eta$:
\begin{eqnarray}
 \eta &=&
\frac{2}{3}\left(120a_6\phi_0^3+24a_4\phi_0\right)^2v_dTZ^{-3}\Lambda^{d-6}\Bigg\{
\nonumber\\
&&
 \frac{6}{d}\frac{1}{(1+w)^5}
 \Big[2(1+w)-\frac{8}{d+2}\Big] - 
 \frac{1}{2d} {\tilde T}^{-5}
 \int_0^1 dy y^{(d+3z)/2-7}
\nonumber\\
&&
 \Bigg[-6 y^2 (d-1) \tilde{T}^2
 \psi_2(x) - 2y^{5/2} \Big(2(8-d)
\nonumber\\
&&
 (1 + w - y) +
 3[(d-14)(1+ w) + 8y] + 27(1+w)  \Big)
\nonumber\\
&&
 \tilde{T} \psi_3(x) - y^{3}
 [2y  + w+1]^2 \psi_4(x) \Bigg]\Bigg\} \; ,
\label{eq:etas}
\end{eqnarray}
where $x=1+\tilde{T}^{-1}(1+w)y^{1/2}$. As before, one identifies a classical contribution, originating from $\omega=0$, and the remaining quantum contribution (involving $\tilde{T}$).

\section{Zero temperature}
\begin{figure}[ht!]
\begin{center}
\includegraphics[width=3.2in]{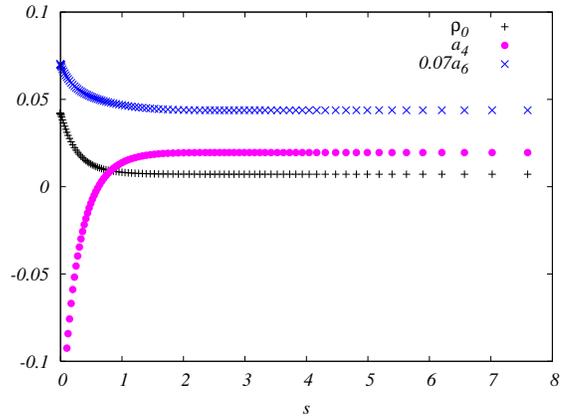}
\caption{(Color online) Solution to the flow equations Eqs.~(\ref{flow_rho_0}, \ref{flow_a4}, \ref{flow_a6}) for $d=2$, $T=0$ and the initial couplings chosen as follows: $a_2(0)=0.004$, $a_4(0)=-0.15$, $a_6(0)=1$, and where $s=-\log{\Lambda/\Lambda_0}$. The parameter $\rho_0$ is renormalized towards lower values, which leads to expanding the region of the phase diagram where no symmetry-breaking occurs. The coupling $a_4$ increases and reaches a positive value (specific to systems exhibiting a second-order transition) in the final part of the flow.  }
\label{flow_d2z3T0}
\end{center}
\end{figure}

\begin{figure}[ht!]
\begin{center}
\includegraphics[width=3.2in]{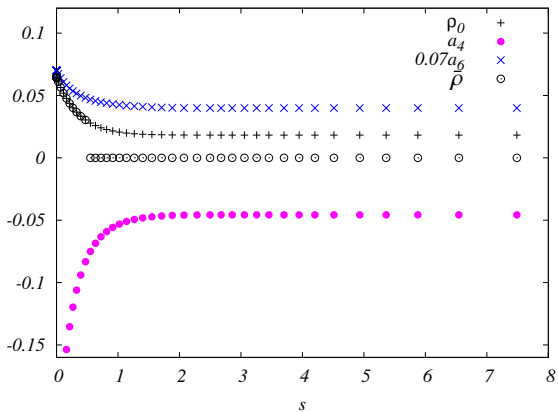}
\caption{(Color online) Solution to the flow equations Eqs.~(\ref{flow_rho_0}, \ref{flow_a4}, \ref{flow_a6}) for $d=2$, $T=0$ and the initial couplings chosen as follows: $a_2(0)=0.0146$, $a_4(0)=-0.25$, $a_6(0)=1$, and where $s=-\log{\Lambda/\Lambda_0}$. The parameter $\bar{\rho}$ corresponding to the global minimum of $U$ exhibits a jump as $U[\rho_0]$ is shifted above $U[\rho=0]$.   }
\label{flow_d2z3T0_dod}
\end{center}
\end{figure}

In this section we present results for the renormalized effective potential at $T=0$. As the critical behavior is governed by a Gaussian fixed point in  this case,\cite{hertz76} we put $\eta=0$. The classical contributions to the threshold functions vanish, while the integrals in their quantum parts can be carried out analytically. For $l_1^{d(Q)}$ one obtains:
\begin{equation}
 l_1^{d(Q)}(w)\to \frac{Z\Lambda^3}{\pi Z_\omega}\frac{1}{1+w}\frac{1}{\frac{1}{2}(d+3)-1}
\label{threshold_11}
\end{equation}
as $T\to 0$. The remaining threshold functions are evaluated by taking consecutive derivatives of Eq.~(\ref{threshold_11}) with respect to $w$. The flow equations are then solved numerically in $d=2$, and $d=3$ for $Z=1$, $Z_\omega=1$, and for different choices of initial couplings corresponding to a first order transition at mean-field level. Namely, we consider the initial value of $a_6=1$, $a_4$ negative, and $a_2$ ranging between small negative and small positive values. For the numerical computations we set the upper cutoff $\Lambda_0=1$. The renormalized phase diagrams are constructed by inspecting the form of the renormalized effective potential (at $\Lambda\to 0$) for the different choices of the initial couplings. The locus of phase transition is identified with the line in the $(a_2, a_4)$ plane, where the global minimum of $U$ becomes nonzero, which happens continuosly or discontinuously as function of $a_2$ or $a_4$. The metastable states (corresponding to local minima of $U$) persist within a region in the $(a_2, a_4)$ plane in the vicinity of the first order transition line.  An example solution of the flow equations in case where the effective potential is renormalized to a form corresponding to a state in the vicinity of a second-order transition, is provided in Fig.~\ref{flow_d2z3T0}. 
In Fig.~\ref{flow_d2z3T0_dod} we present a solution for a situation where the absolute minimum of $U$ becomes zero in course of the flow, while the metastable states persist for $\Lambda\to 0$. Therefore, the average order parameter exhibits a discontinuity as function of $\Lambda$. 

\begin{figure}[ht!]
\begin{center}
\includegraphics[width=3.2in]{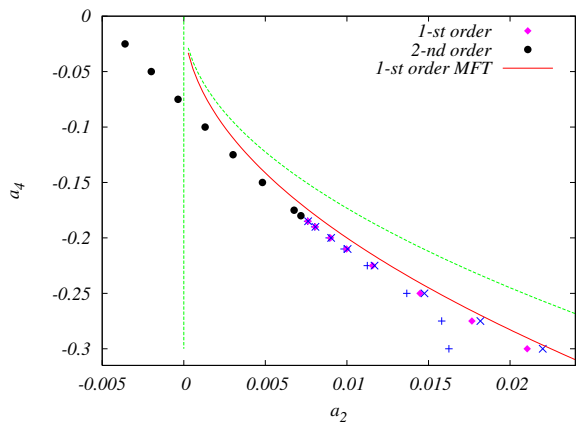}
\caption{(Color online) Zero temperature phase diagram of the renormalized $\phi^6$ model in $d=2$. The renormalized tricritical point is shifted towards negative $a_4$. The dashed lines and crosses mark boundaries of the regions where metastable configurations are present in bare and renormalized theory, respectively.       }
\label{fig:plot_d2z3}
\end{center}
\end{figure}

\begin{figure}[ht!]
\begin{center}
\includegraphics[width=3.2in]{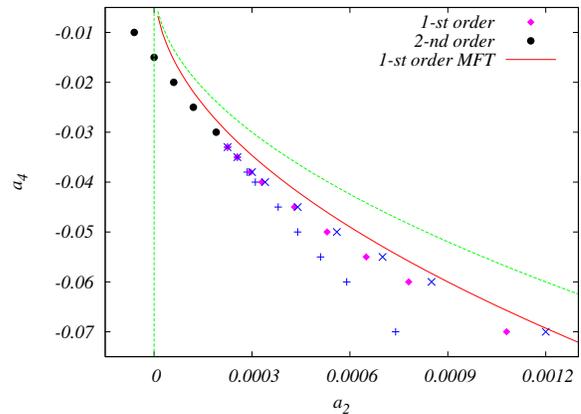}
\caption{(Color online) Zero temperature phase diagram of the renormalized $\phi^6$ model in $d=3$. The renormalized tricritical point is shifted towards negative $a_4$. The dashed lines and crosses mark boundaries of the regions where metastable configurations are present in bare and renormalized theory, respectively.       }
\label{fig:plot_d3z3}
\end{center}
\end{figure}

The results for the renormalized phase diagrams in the variables $(a_2, a_4)$ for $d=2$ and $d=3$ are presented in Figs.~\ref{fig:plot_d2z3}, \ref{fig:plot_d3z3}.

As anticipated, the region corresponding to the disordered phase in the phase diagram is extended in course of renormalization. One also finds that upon including order parameter fluctuations, the tricritical point is shifted towards negative values of $a_4$. The magnitude of this shift is nearly one order of magnitude larger in the case $d=2$, where fluctuations are more pronounced. We also note that the renormalization of the system parameters occurs at the beginning of the flow only. For the present case $T=0$ it is caused exclusively by quantum fluctuations. A consequence of the results in Figs.~\ref{fig:plot_d2z3}, \ref{fig:plot_d3z3} is that there exists a regime of parameters, for which the quantum phase transition is first order at mean-field level (within the bare theory), and turns continuous when fluctuations are accounted for. This change of the quantum phase transition's character does not occur in the other direction within the framework applied here. 

In the subsequent section, we analyze the system at  finite temperatures and argue that the tendency towards second order scenario is even stronger when thermal fluctuations are present. Therefore two interesting possibilities are conceivable - either the transition is second order for $T\geq 0$, or it becomes of second order for temperatures exceeding a tricritical temperature $T^{tri}$. Of special interest is the case $T^{tri}=0$, where the finite $T$ second order phase transition line terminates at $T=0$ at a quantum tricritical point.  

\section{Finite temperatures}
In this section we discuss renormalization of the effective potential $U[\phi]$ in case of finite temperatures, where both quantum and thermal fluctuations are present. Conceptually, the analysis proceeds along the same lines as for $T=0$, however the threshold functions now contain both  thermal and quantum parts and the integrals they involve cannot be carried out analytically. Moreover, the anomalous dimension $\eta$ has to be accounted for, as we encounter non-Gaussian critical behavior in the vicinity of the transition line. We also note, that even slightly off criticality, the presence of anomalous dimension influences the results for the phase diagram, in particular in $d=2$. This happens because $\eta$ attains nonzero values at the intermediate stages of the flow, and vanishes only in the infrared limit. As a result, the factor $Z$ acquires finite, though not negligibly small renormalization. As we checked by explicit calculations, a truncation with $Z=1$, which breaks down only in the immediate vicinity of a second-order transition, is not sufficient to determine the order of the transition. Namely, at least in $d=2$, it gives a very weakly first order transition in the regime where the truncation retaining the flow of $\eta$ predicts a second order scenario. As we also checked, the present truncation taking into account the flow of $Z$ reproduces the results of Sec. IV in the limit $T\to 0$. The relative difference in the numerical results for the points exhibited in Figs 5,6  using the two different truncations is below 1\%. 

\begin{figure}[ht!]
\begin{center}
\includegraphics[width=3.2in]{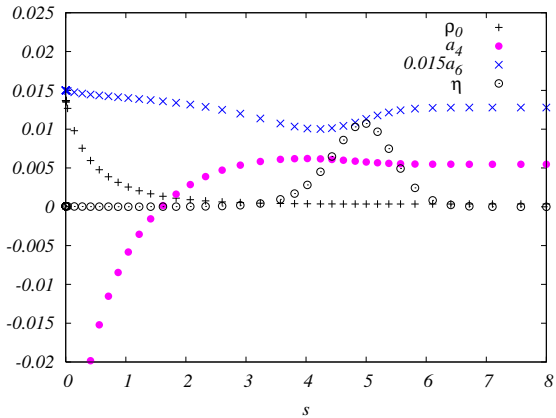}
\caption{(Color online) Solution to the flow equations Eq.~(\ref{flow_rho_0}, \ref{flow_a4}, \ref{flow_a6}, \ref{eq:etas}) for $d=3$, $T=0.08$ and the initial couplings chosen as follows: $a_2(0)=0.00049$, $a_4(0)=-0.05$, $a_6(0)=1$, and where $s=-\log{\Lambda/\Lambda_0}$. This choice of parameters corresponds to the phase with broken symmetry, but close to the transition line (within the renormalized theory). The coupling $a_4$ increases and reaches a positive value in the final part of the flow. The anomalous dimension $\eta$ attains nonzero values in the intermediate stages of the flow, although the system is slightly off criticality, and vanishes in the infrared limit. The system exhibits a second order phase transition upon varying $a_2$, although for the same choice of $a_4(0)$, $a_6(0)$, the transition is of first order at $T=0$ - see Fig.~\ref{fig:plot_d3z3}.} 
\label{flow_d3z3T0.08}
\end{center}
\end{figure}

 Therefore, we analyze the set of flow equations Eqs.~(\ref{flow_rho_0}, \ref{flow_a4}, \ref{flow_a6}, \ref{eq:etas}) with the aim of determining the phase diagrams in the variables $(a_2, a_4)$ for different temperatures. In a subsequent step we reinterpret our results by considering fixed (initial) $a_4<0$ and plotting phase diagrams in the variables $(a_2,T)$. As before, we set the upper cutoff $\Lambda_0=1$, for the initial condition we choose $a_6=1$, $Z=1$, and consider negative initial values of $a_4$, and $a_2$ ranging between small negative and small positive values. The factor $Z_\omega$ is set equal to unity. An example solution to the flow equations is given in Fig.~\ref{flow_d3z3T0.08}.

In the solution illustrated in Fig.~\ref{flow_d3z3T0.08}, the choice of $a_4$ corresponds to a situation where, despite the fact that $a_4$ is negative,   one encounters a second order transition upon varying $a_2$. Moreover, for this value of $a_4$, the transition is of first order at $T=0$, which can be read off from Fig.~\ref{fig:plot_d3z3}. The initial value of $a_2$ was chosen so that the system is slightly separated from the transition, but $\eta$ attains nonzero values at an intermediate stage of the flow. 

\begin{figure}[ht!]
\begin{center}
\includegraphics[width=3.0in]{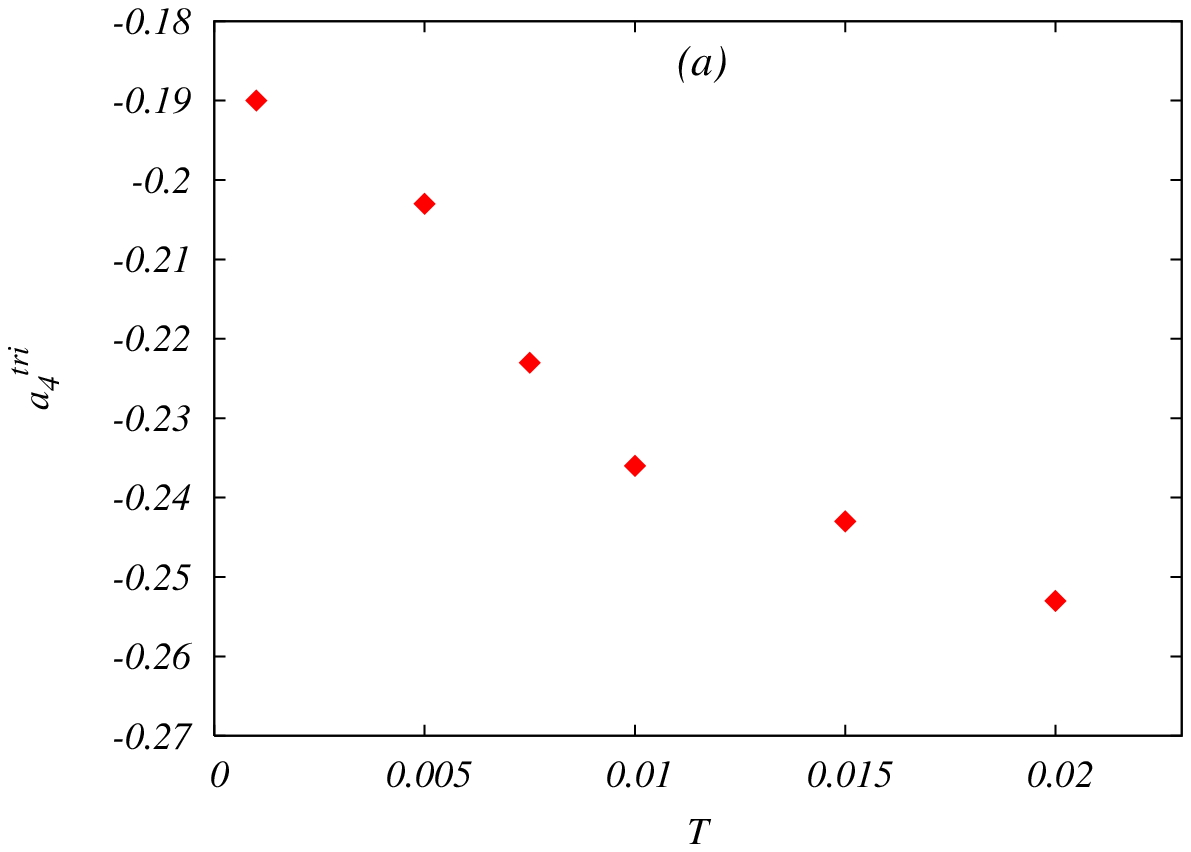}
\includegraphics[width=3.0in]{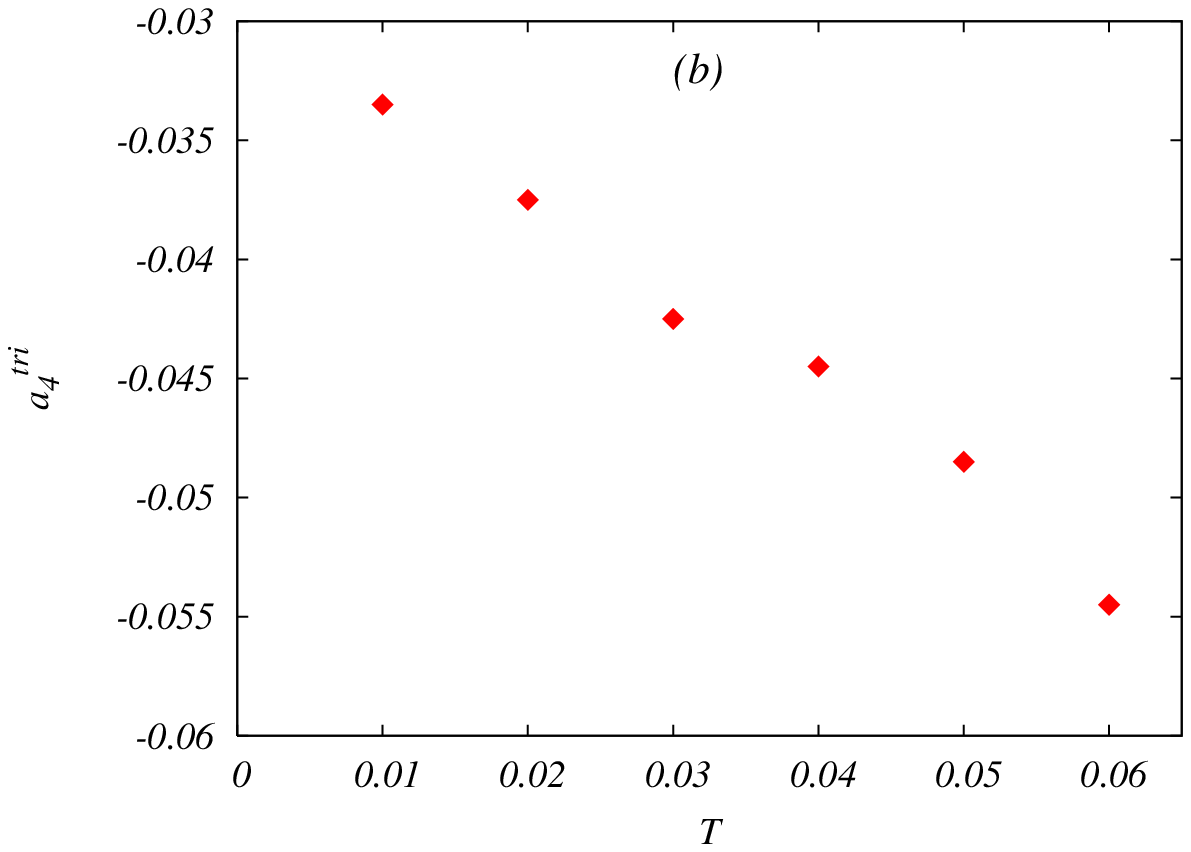}
\caption{(Color online) The function $a_4^{tri}(T)$ computed from Eqs.~(\ref{flow_rho_0}, \ref{flow_a4}, \ref{flow_a6}, \ref{eq:etas}) for (a) $d=2$ and (b) $d=3$. For $a_4>a_4^{tri}$ the system exhibits a second order transition upon varying $a_2$, while for $a_4<a_4^{tri}$ the transition is of first order.}
\label{a_4_tri}
\end{center}
\end{figure}

By performing such an analysis for different values of $T$, $a_4$, and $a_2$, we obtain (for different fixed $T$), phase diagrams analogous to those plotted in Figs~\ref{fig:plot_d2z3}, \ref{fig:plot_d3z3} for $T=0$. Qualitatively these phase diagrams are similar, however the position of the renormalized tricritical point $a_4^{tri}$ depends on temperature. We find that both for $d=2$ and $d=3$ $a_4^{tri}(T)$ is a decreasing function, and therefore the tendency towards a second order transition is enhanced upon increasing $T$. In Fig.~\ref{a_4_tri}, we plot the evaluated $a_4^{tri}(T)$ for $d=2$ and $d=3$. The quantity $a_4^{tri}(T)$ may in general be a complicated function of $a_6$ as well as the upper cutoff $\Lambda_0$, and for more sophisticated truncations also other parameters characterizing the effective action. Therefore, its precise form depends on all these parameters. We believe however, that the function would remain monotonously decreasing also with better truncations, provided the qualitative structure of the phase diagram for fixed $T$ is as in Fig. 1. This could change for example after including a $\phi^8$ term and allowing for negative $a_6$, which however corresponds to a physical situation different to the one studied in this paper. 

\begin{figure}[ht!]
\begin{center}
\includegraphics[width=3.0in]{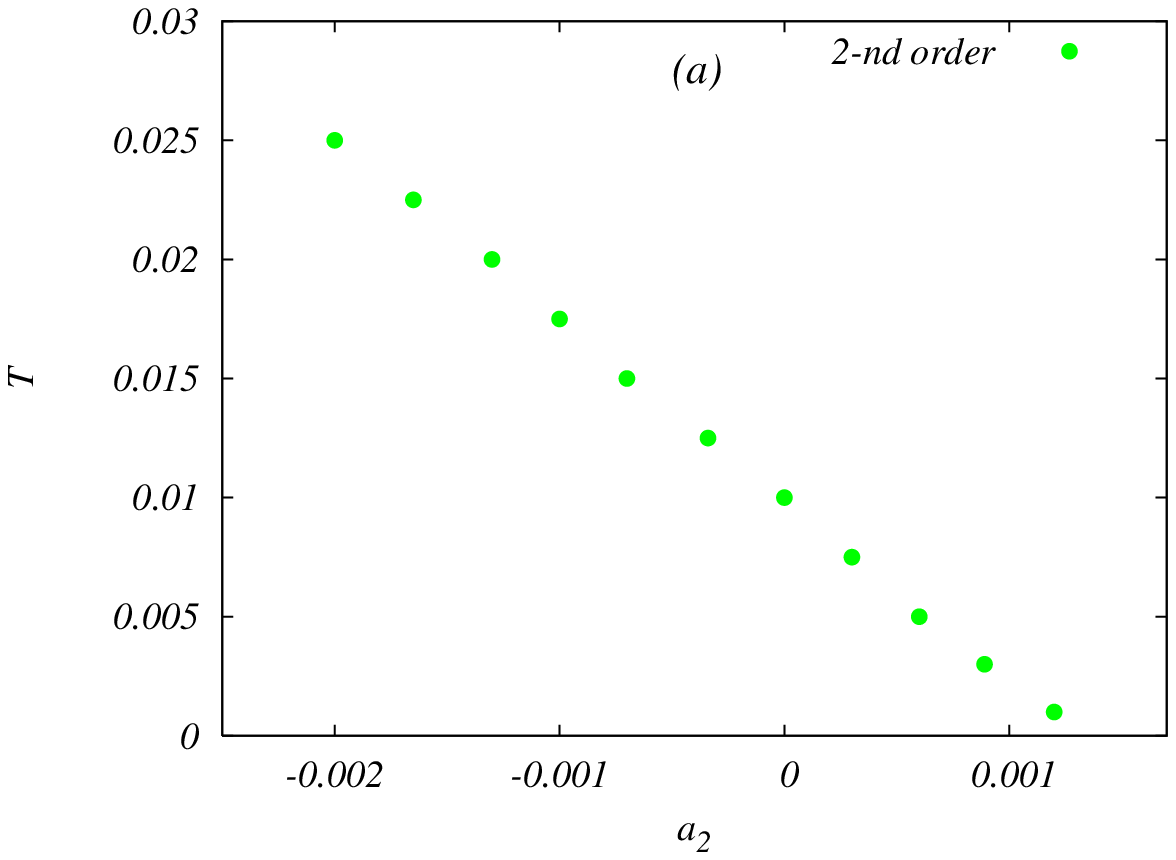}
\includegraphics[width=3.0in]{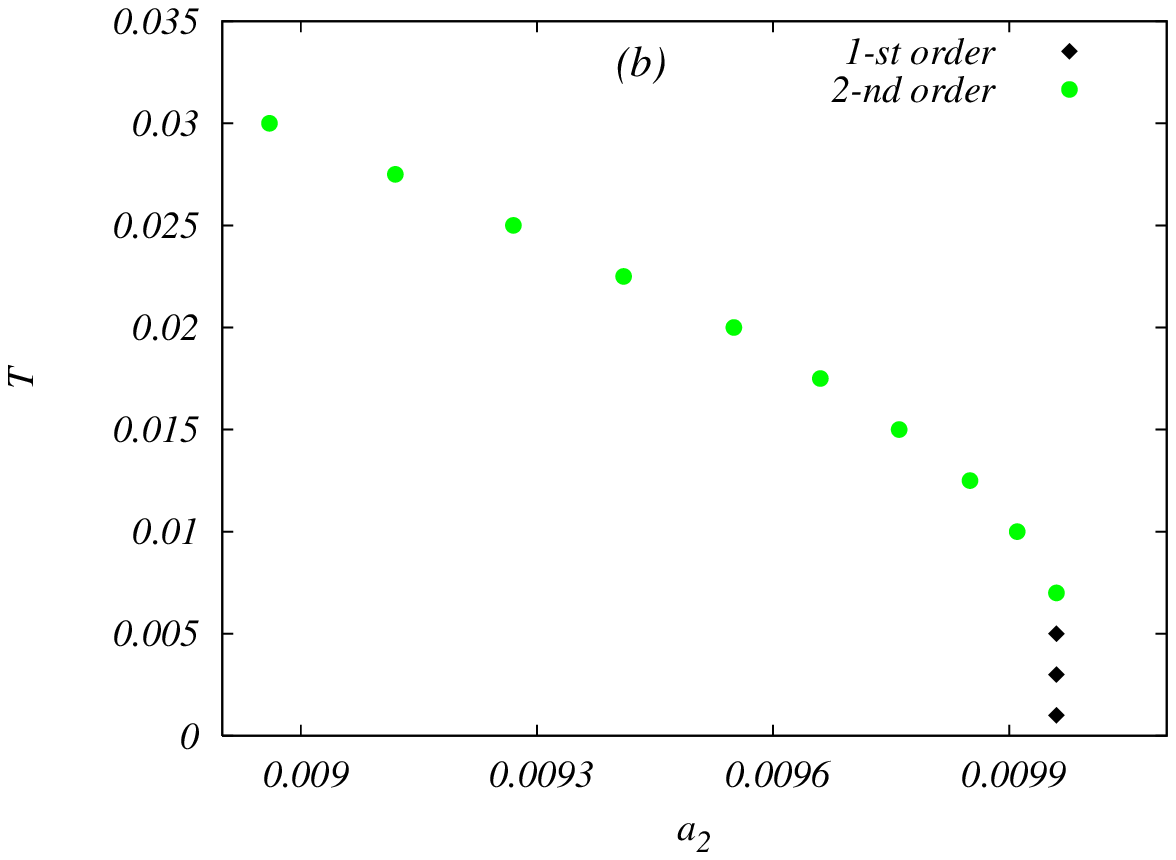}
\caption{(Color online) Transition temperature as function of the control parameter $a_2$ as computed from Eqs.~(\ref{flow_rho_0}, \ref{flow_a4}, \ref{flow_a6}, \ref{eq:etas}) in $d=2$. The ordered phase is situated below the transition line. The plots correspond to (a) $a_4=-0.10$, and (b) $a_4=-0.21$. The transition line (a) terminates at a quantum critical point at $T=0$, while in case (b) the quantum phase transition is of first order, and a tricritical point is present at $T\approx 0.006$.     }
\label{phase_diag_d2_Temp}
\end{center}
\end{figure}

\begin{figure}[ht!]
\begin{center}
\includegraphics[width=3.0in]{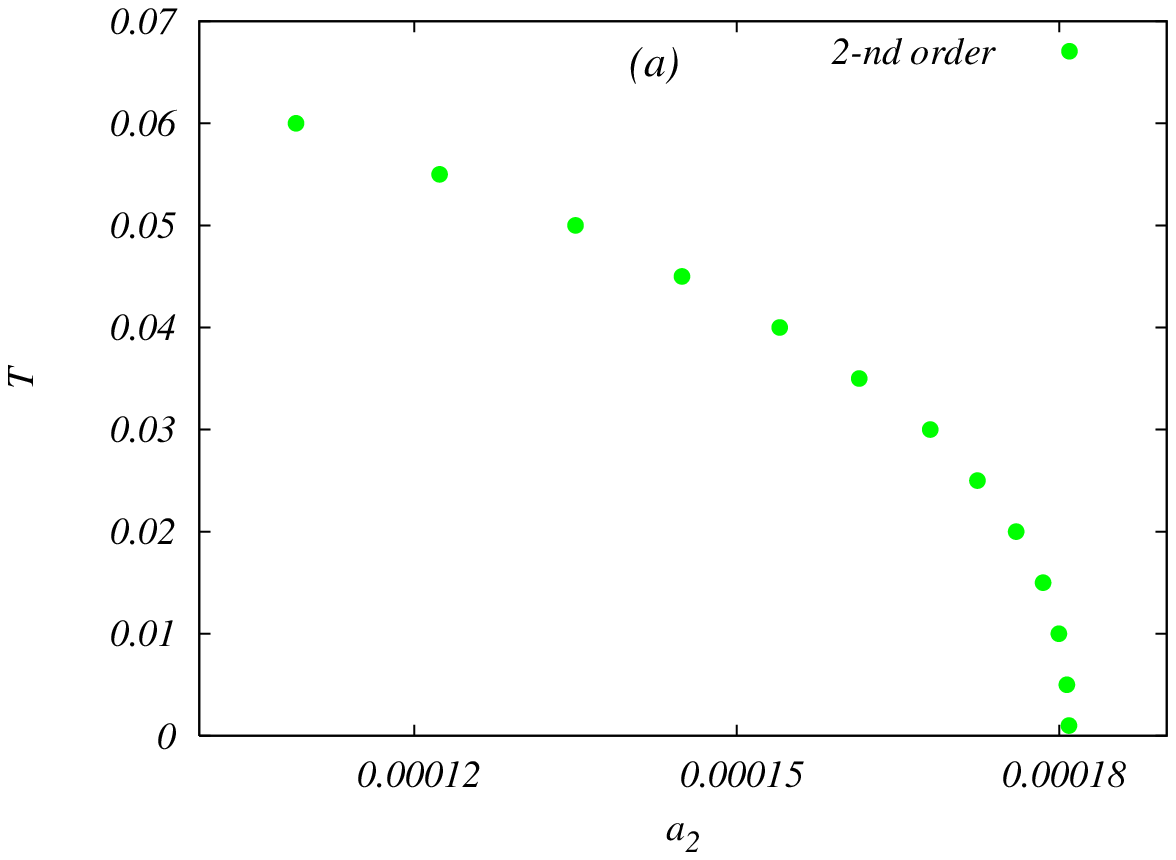}
\includegraphics[width=3.0in]{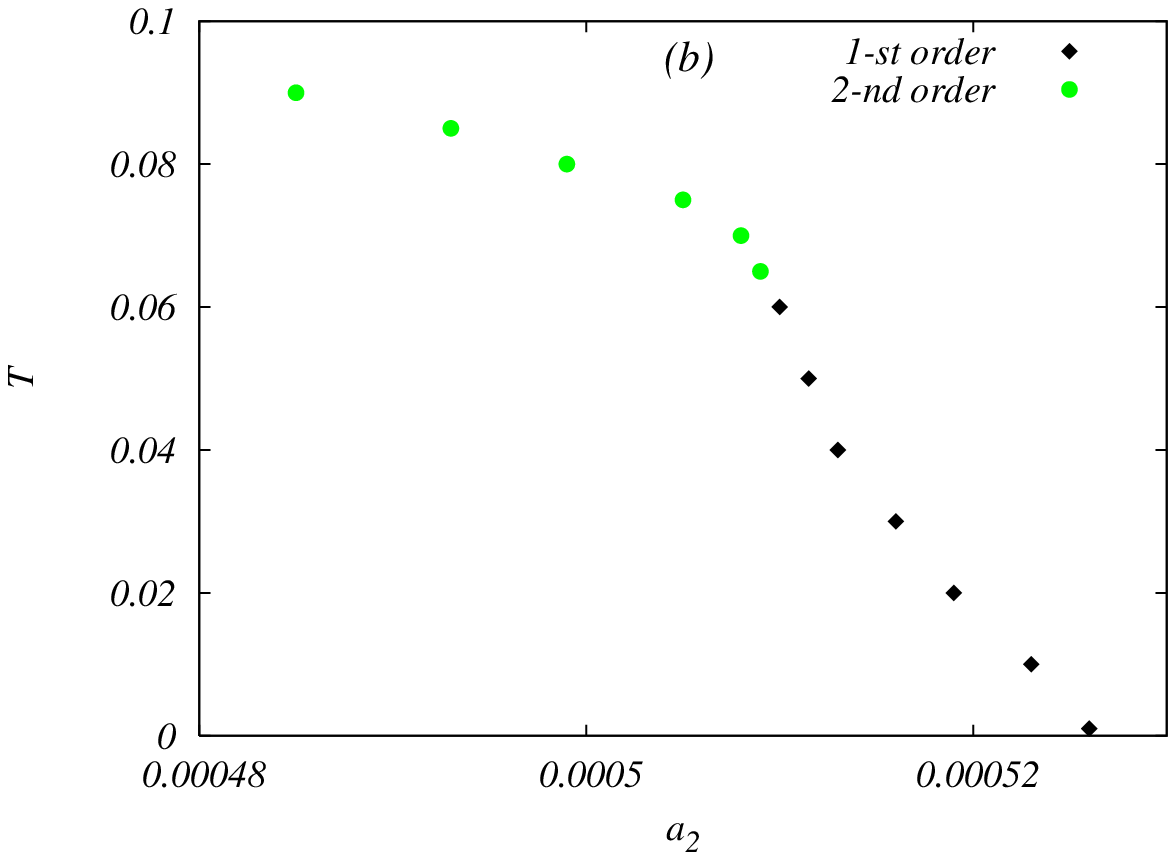}
\caption{(Color online) Transition temperature as function of the control parameter $a_2$ as computed from Eqs.~(\ref{flow_rho_0}, \ref{flow_a4}, \ref{flow_a6}, \ref{eq:etas}) in $d=3$. The ordered phase is situated below the transition line. The plots correspond to (a) $a_4=-0.03$, and (b) $a_4=-0.05$. The transition line (a) terminates at a quantum critical point at $T=0$, while in case (b) the quantum phase transition is of first order, and a tricritical point is present at $T\approx 0.05$.     }
\label{phase_diag_d3_Temp}
\end{center}
\end{figure}

We proceed by evaluating the phase diagrams in the variables $(a_2, T)$ for a fixed (initial) coupling $a_4$. For this aim one solves the flow equations for different $T$ and $a_2$, inspects the resulting renormalized $U$ and identifies the relevant global minimum.  We consider two possibilities: $a_4^{tri}(T=0)<a_4<0$, in which case the transition is of second order for $T\geq 0$; and $a_4<a_4^{tri}(T=0)$, where the quantum phase transition is first order and one encounters a tricritical point at some $T^{tri}>0$. The results are summarized in Figs~\ref{phase_diag_d2_Temp}, \ref{phase_diag_d3_Temp}.
In cases where the transition is first order for low temperatures we find approximately linear behavior of the phase boundary in the limit $T\to 0$ both in $d=2$ and $d=3$. However, in $d=2$ the slope of the transition line is infinite (or very large), unlike in $d=3$. An important conclusion is that a first order transition occurs for $T<T^{tri}$ in the considered situations. We do not expect the shapes of the first order transition lines to be universal and therefore, in our opinion, it is not quite justified to compare the results of the simple $\phi^6$ model to other calculations without first assessing the validity of this approach in the considered context. One may note however, that mean field calculations of the first order phase boundaries in two dimensional systems exhibiting discrete symmetry breaking Fermi surface deformations yield predictions qualitatively similar to our results \cite{yamase05} (very steep, linear behavior of the phase boundary for $T\to 0$). Our expectation is that the results of Fig.~\ref{phase_diag_d2_Temp} (b) may apply to the system considered in Ref.~\onlinecite{yamase05} for nonzero uniform fermionic repulsion term $u$, which is required to assure that the factor $a_6$ in the corresponding Landau expansion is positive for temperatures reaching down to zero. The other requirement is that the tricritical point is located close to $T=0$ (as in Fig. 4 (b) in the abovementioned work). When the results for $d=3$ are compared to data from experiments on compounds exhibiting ferromagnetic properties (see e.g. Refs.~\onlinecite{pfleiderer97, pfleiderer04}), one finds qualitative agreement in that the first order transition is located at lower $T$, reaching down to $T=0$, while a second order transition occurs at slightly higher $T$. However, another mechanism driving the transition first order is present for the case of ferromagnets, and the corresponding free energy functional involves terms logarithmic in $\phi$.\cite{belitz99} This fact is not accounted for within the present approach.   As regards to the situations where a quantum critical point is realized, we recover approximately linear behavior of the phase boundary in $d=2$, as dictated by the Hertz-Millis theory. The shapes of the phase boundaries in such case for $d=3$ are discussed in more detail in subsection V B.

\subsection{Order parameter exponents}
In this subsection we analyze the behavior of the order parameter upon approaching the phase transition line along isotherms for which different situations occur. We also compute the magnitude of the order parameter jump when approaching the tricritical point along the coexistence curve and extract the corresponding critical exponents. For the purpose of this subsection we put $a_4=-0.05$ and $a_4=-0.21$ for $d=3$ and $d=2$, respectively. This means we focus on the situations depicted in Fig.~\ref{phase_diag_d2_Temp} (b) and Fig.~\ref{phase_diag_d3_Temp} (b). An investigation of the critical exponents characterizing the second-order transition in $d=2,3$ in the present context is contained in Ref.~\onlinecite{jakubczyk08}, where the results are also compared to calculations involving more sophisticated functional RG truncations \cite{canet03, ballhausen04} and exact values from the Onsager solution to the Ising model. 

\begin{figure}[ht!]
\begin{center}
\includegraphics[width=3.0in]{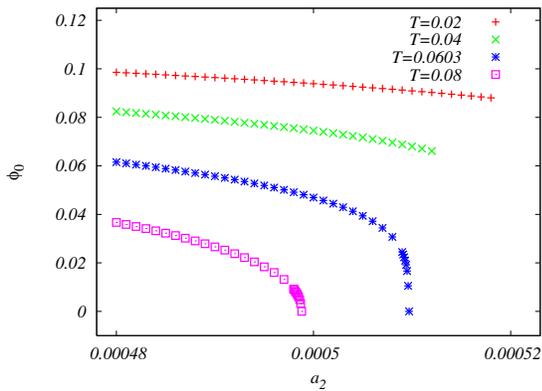}
\caption{(Color online) The order parameter $\phi_0$ plotted vs $a_2$ as the transition line is approached along the isotherms $T=0.02$, $T=0.04$, $T=0.0603$ and $T=0.08$. For $T=0.02$, $T=0.04$, where the transition is first order, a discontinuity of $\phi_0$ occurs at the phase boundary. For $T=0.0603$, corresponding to the tricritical isotherm, the order parameter follows the power law $\phi_0\sim (a_2^{tri}-a_2)^{\beta_{tri}}$, where $\beta_{tri}\approx 0.25$. Along the isotherm, $T=0.08$, one observes a power law $\phi_0\sim (a_2^{cr}-a_2)^{\beta}$, where $\beta=0.5$. Mean-field behavior is observed because the plotted points are outside the tiny non-gaussinan regime.}
\label{Order_parameters_d3}
\end{center}
\end{figure}

For $d=3$ we computed the order parameter, which is given by $\lim_{\Lambda\to 0}\phi_0$ ($\phi_0$ being the global minimum of $U[\phi]$) as a function of $a_2$, approaching the phase transition line along the isotherms $T=0.02$, $T=0.04$, $T=0.0603$ and $T=0.08$. The results are plotted in Fig.~\ref{Order_parameters_d3}.
 For $T=0.02$ and $T=0.04$ the order parameter exhibits a discontinuity, as the minimum at $\phi=0$ becomes the global one when the phase boundary is reached. The order parameter jump vanishes at the tricritical isotherm $T\approx 0.0603$, along which the order parameter follows the power law $\phi_0\sim (a_2^{tri}-a_2)^{\beta_{tri}}$, where $\beta_{tri}\approx 0.25$, as anticipated within mean-field theory.\cite{lawrie84} To recover the expected logarithmic corrections to the leading power law behavior requires going very close to the tricritical point, which is not achieved here. Along the isotherm, $T=0.08$, in the scale exhibited in Fig.~\ref{Order_parameters_d3},    one observes a power law $\phi_0\sim (a_2^{cr}-a_2)^{\beta}$, where $\beta=0.5$ is the mean-field value. In fact, a different behavior, with $\beta \approx 0.31$ is expected to occur sufficiently close to the critical line. This is not visible in Fig.~\ref{Order_parameters_d3}  because the plotted points are still outside the tiny region where the Ginzburg criterion is violated. The crossover to the non-gaussian behavior as well as sizes of the truly critical region as a function of temperature $T$ were studied in detail in Ref.~\onlinecite{jakubczyk08} both in $d=2$ and $d=3$. 

\begin{figure}[ht!]
\begin{center}
\includegraphics[width=3.0in]{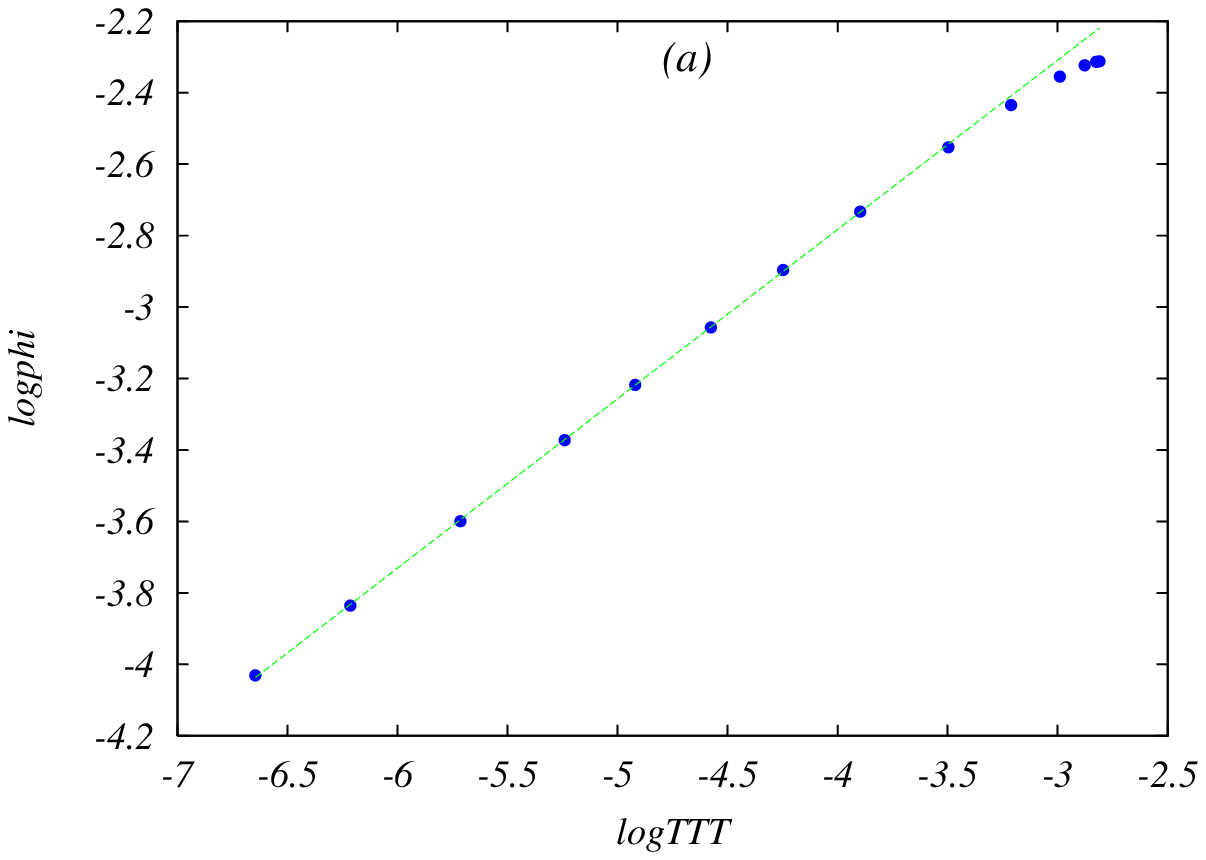}
\includegraphics[width=3.0in]{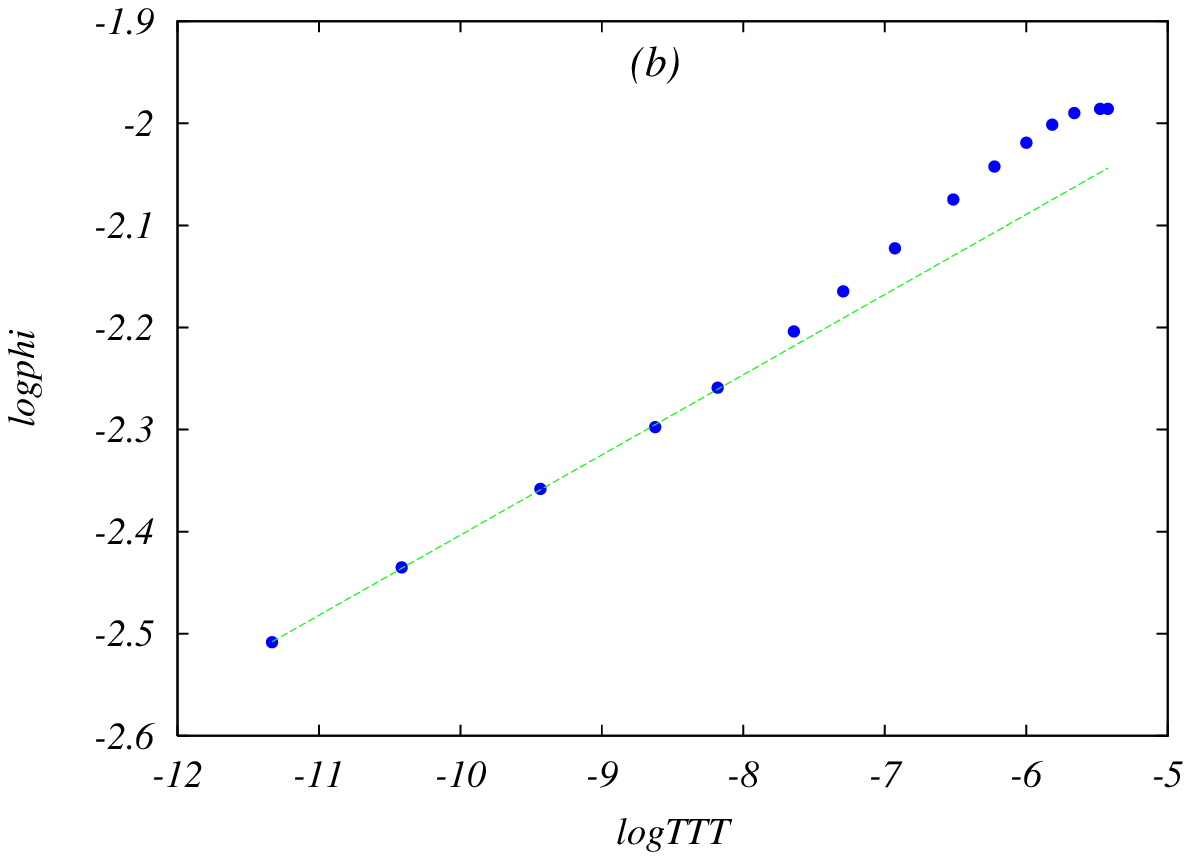}
\caption{(Color online) The order parameter jump plotted vs $(T^{tri}-T)$ as the tricritical point is approached along the coexistence line for (a) $d=3$,  $a_4=-0.05$, (b) $d=2$, $a_4=-0.21$. In the vicinity of the tricritical point the computed order parameter jumps follow a power law $\phi_0\sim (T^{tri}-T)^{\mu}$. For $d=3$ one finds $\mu\approx 0.48$, being close to the mean-field value. For $d=2$ $\mu\approx 0.079$, deviating from mean field behavior.}
\label{ord_par_jumps}
\end{center}
\end{figure}

We proceed by analysing the order parameter jump upon varying temperature towards $T^{tri}$ along the coexistence curve. The results for $d=2,3$ are plotted in Fig.~\ref{ord_par_jumps}.
In $d=3$ the jump of the order parameter follows a power law $\phi_0\sim (T^{tri}-T)^{\mu}$ with an exponent $\mu\approx 0.48$ close to the mean field value $0.5$. This behavior breaks down at lower temperatures, where the system's behavior is no longer expected to be universal. For $d=2$ a power law with a non mean field exponent $\mu\approx 0.079$ is observed. According to our knowledge, the classical tricritical behavior in $d=2$ was not addressed within the functional RG framework. The obtained result compares relatively well to the accurate value $\mu\approx 0.094$, \cite{lawrie84} bearing in mind the simplicity of the applied approximation. A better estimate of $\mu$ would require a more sophisticated truncation of functional RG. This, and also other aspects of classical tricritical behavior in $d=2$ remain to be addressed in a separate study.

\subsection{Crossover of the shift exponent}
\begin{figure}[ht!]
\begin{center}
\includegraphics[width=3.0in]{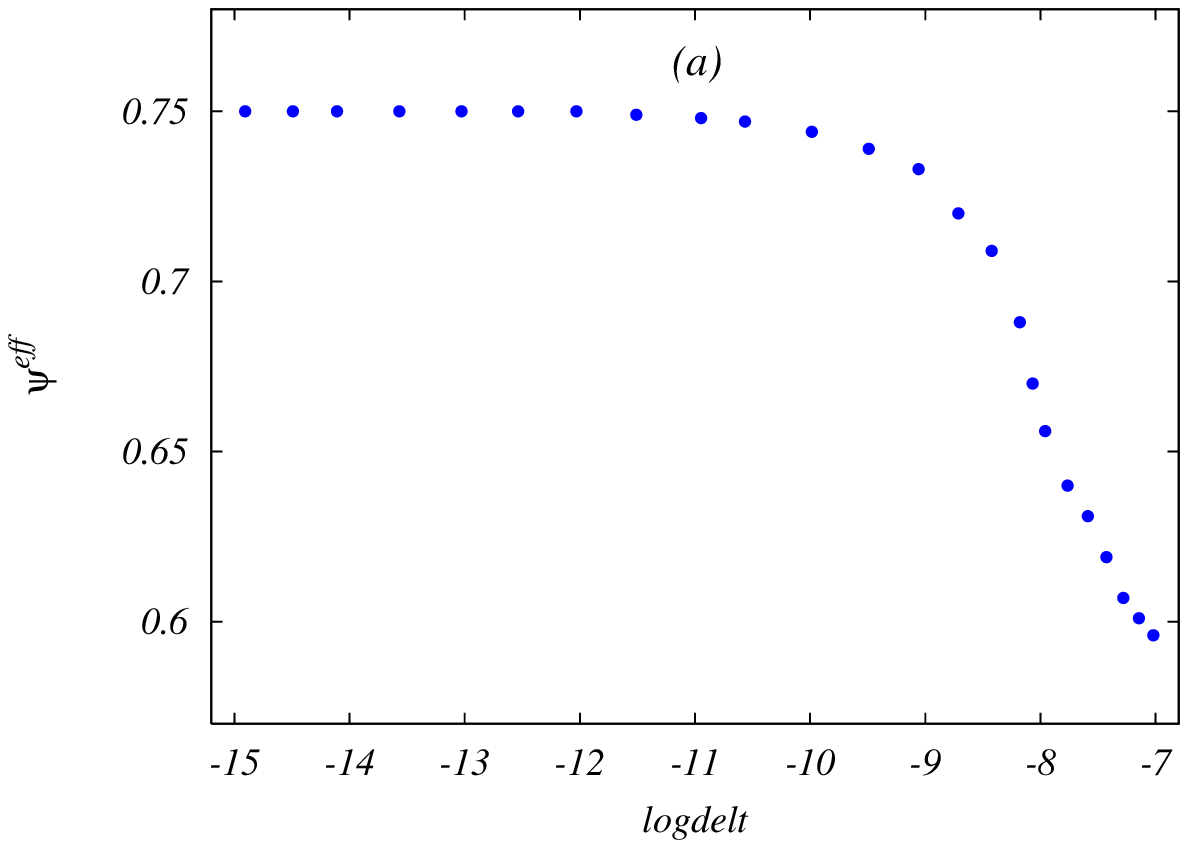}
\includegraphics[width=3.0in]{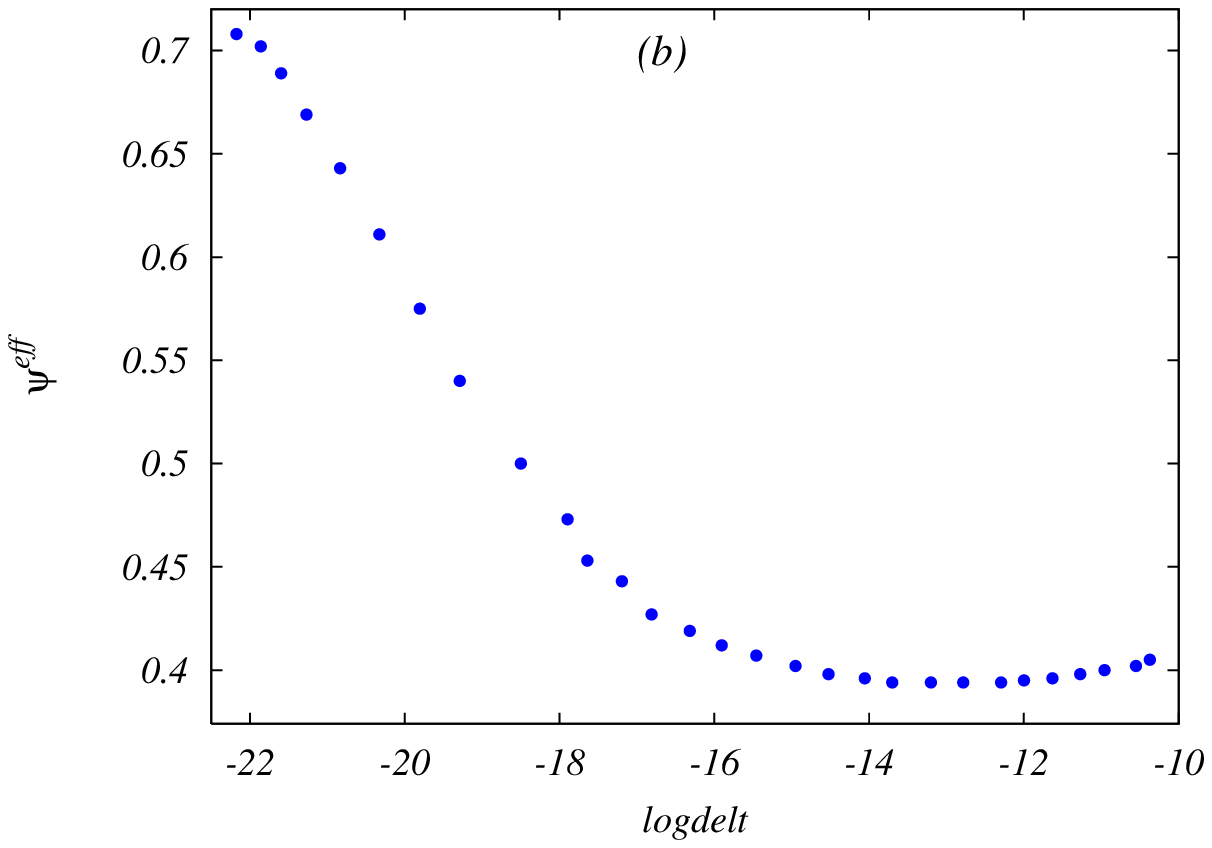}
\caption{(Color online) The effective shift exponent $\psi^{eff}$ plotted as a function of $|a_2-a_2^{(0)}|$ for (a) $a_4=-0.01$, and (b) $a_4=-0.0302$. Case (b) corresponds to close vicinity of $a_4^{tri}(T=0)$. The transition line exhibits scaling behavior as dictated by Hertz-Millis theory for the case (a). For case (b) a scaling regime with the shift exponent $\psi\approx 3/8$, specific to quantum tricriticality is observed (see the main text), and the Hertz-Millis scaling persists only in the narrow vicinity of $a_2^{(0)}$, which is not visible in the picture scale.     }
\label{psi_eff}
\end{center}
\end{figure}

Here we present results regarding the phase boundaries in $d=3$ in the case, where the transition at $T>0$ is second order. We consider two choices of the initial coupling $a_4$: $a_4=-0.01$, which is separated from $a_4^{tri}(T=0)$, and $a^4=-0.0302$, a value in the proximity of $a_4^{tri}(T=0)$. We perform a careful numerical analysis of the shape of the transition lines over a relatively wide range of temperatures and extract the effective shift exponent $\psi^{eff}$ as a function of $|a_2-a_2^{(0)}|$. The quantity $\psi^{eff}$ is calculated by fitting a power law to three neighboring points of the transition line, which is computed numerically. Scaling behavior of the phase boundary shape occurs if  $\psi^{eff}$ remains constant over a wide range of $|a_2-a_2^{(0)}|$. Our findings are summarized in Fig.~\ref{psi_eff}.
From Fig.~\ref{psi_eff} (a) we read off the standard value of the shift exponent $\psi=3/4$, as dictated by Hertz-Millis theory. However, for case (b) scaling behavior with $\psi\approx 3/8$ is observed. This is related to the proximity of the chosen coupling $a_4$ to the tricritical value  $a_4^{tri}(T=0)$. We also performed analogous computations for other values of $a_4$ between $-0.01$ and $-0.03027$, yielding the observation, that the Hertz-Millis scaling region shrinks as $a_4$ approaches $a_4^{tri}(T=0)$, and the other scaling regime forms sufficiently close to $a_4^{tri}(T=0)$. As $a_4\to a_4^{tri}(T=0)$, the value of $\psi$ corresponding to the latter scaling platoe approaches $3/8$.  In the subsequent section we provide an analytic argument to further clarify the emergent picture and justify the obtained value $\psi^{tri}=3/8$. 

\section{Scaling analysis}
In this section we show how the obtained values of the shift exponents $\psi^{HM}$ and $\psi^{tri}$ can be recovered by invoking a scaling relation for the free energy in the vicinity of the quantum critical point (see e.g. Ref.~\onlinecite{sachdev_book}). The analysis largely follows Ref.~\onlinecite{belitz_review05}, where $\psi^{HM}$ was derived, and is restricted to $d>2$. At $d=2$ additional logarithmic corrections occur, which are not accounted for here. The homogeneity relation for the free energy density
\begin{equation}
 f(a_2,T, h, a_n)=b^{-d-z}f(a_2 b^{1/\nu}, Tb^z, hb^{y_h}, a_nb^{[a_n]})
\label{sc_rel}
\end{equation}
is written including the magnetic field $h$ and the most relevant interaction coupling $a_n$. Although the scaling dimension $[a_n]<0$, it influences the critical behavior because $a_n$ acts as a so-called dangerously irrelevant variable. \cite{goldenfeld_book} Now we differentiate Eq.~(\ref{sc_rel}) with respect to $h$ and put $h=0$, thus obtaining a homogeneity relation for the average order parameter $\phi_0$ at $h=0$. In the next step we assume, that the dependence of $\phi_0$ on $a_2$ and $a_n$ occurs only via the ratio $a_2/a_n$. This holds within mean field theory, and should not change the value of the shift exponent provided this exponent is the same for the transition and Ginzburg lines in the vicinity of the quantum critical point. This issue was investigated in Ref.~\onlinecite{jakubczyk08}, where the size of the truly critical region was analyzed yielding the conclusion that the Ginzburg line provides a very accurate estimate of $T_c$ in $d=3$. Therefore we write
\begin{equation}
 \phi_0(a_2/a_n,T)=b^{-d-z+y_h}\phi_0((a_2/a_n)b^{1/\nu-[a_n]}, Tb^z)\;.
\label{sc_phi}
\end{equation}
By choosing the scaling factor $b$ such that the first argument of $\phi_0$ on the RHS of Eq.~(\ref{sc_phi}) becomes a constant, we obtain 
\begin{equation}
 \phi_0(a_2/a_n,T)=(a_n/a_2)^{-\kappa}\phi_0(1, T(a_n/a_2)^{\nu z/(1-\nu [a_n])})\;,
\end{equation}
with $\kappa=\nu (d+z-y_h)/(1-\nu [a_n])$. Now we demand that the order parameter $\phi_0$ vanishes along the transition line $a_2(T)$, that is $\phi_0(a_2(T)/a_n,T)=0$ for any $T$. This yields $T(1/a_2(T))^{\nu z/(1-\nu [a_n])}=const$, and therefore the $\psi$ exponent is identified as 
\begin{equation}
 \psi=\frac{\nu z}{1-\nu [a_n]}\;.
\end{equation}
Once again we now take advantage of the mean field assumption identifying the transition and Ginzburg lines and replace the exponent $\nu$ with its mean field value $\nu^{MFT}=1/2$. Again, this step is justified by invoking the results of Ref.~\onlinecite{jakubczyk08} and noting that it suffices to consider a region of the phase diagram off true criticality, in which mean field scaling holds.  This yields 
\begin{equation}
\psi=\frac{z}{2-[a_n]}\;.
\label{gen_psi}
\end{equation}
By substituting $n=4$ and $[a_4]=4-(d+z)$, we recover the formula $\psi=z/(d+z-2)$, derived by Millis. \cite{millis93} For $d=z=3$ we obtain $\psi=3/4$. However, if the most relevant coupling corresponds to $n=6$, the result is 
\begin{equation}
 \psi=z/((2(d+z)-4)\;, 
\end{equation}
which for $d=z=3$ yields $\psi=\psi^{tri}=3/8$, as obtained numerically in the previous section. 

The corresponding values of the shift exponents can also be extracted from Eq.~(\ref{gen_psi}) for $z=2$, which is not covered by our study in Sec. III-V. For this case one obtains $\psi^{HM}=2/3$ and $\psi^{tri}=1/3$.

Let us also observe, that Eq.~(\ref{gen_psi}), viewed as a function of $[a_n]$, generates a sequence of conceivable shift exponents. For example for $d=z=3$ the sequence is $\psi_n=3/(2n-4)$ for $n=4,6,8,...$. Monotonous decreasing of this sequence is a manifestation of the general fact, that fluctuations tend to suppress the critical temperature. Indeed, as $n$ grows, fluctuation effects get less relevant, $\psi$ decreases, and the $T_c$ line becomes steeper. 

Also note, that for $d\to 2^+$, $\psi$ equals 1 for all $n$ and arbitrary $z$. This follows directly from Eq.~(\ref{gen_psi}). Therefore (up to the neglected logarithmic terms), the behavior of the $T_c$ line in two dimensions is always expected to be linear. 

The analysis performed here and supported by the numerical results of Sec. V provides a clear physical picture of the system under study. Consider that in addition to the UV cutoff $\Lambda_0$ there is another scale $\Lambda^{tri}$  present. This scale is set by $(a_4-a_4^{tri})/a_6$. Once $\Lambda^{tri}>\Lambda_0$, which always holds for $a_6=0$, one recovers Hertz-Millis scaling of $\psi$ for $|a_2-a_2^{(0)}|\ll \Lambda_0$. If $\Lambda^{tri}$ is reduced below $\Lambda_0$, one still observes Hertz-Millis behavior as long as the two scales don't become well separated, the only difference being that scaling now occurs for  $|a_2-a_2^{(0)}|\ll \Lambda^{tri}$. Once $\Lambda^{tri}\ll\Lambda_0$, two scaling regimes are present: for  
$|a_2-a_2^{(0)}|\ll \Lambda^{tri}$ the Hertz-Millis scenario persists, while for $\Lambda^{tri}\ll|a_2-a_2^{(0)}|\ll \Lambda_0$ scaling behavior with $\psi^{tri}=3/8$ occurs. If $\Lambda^{tri}= 0$, the quantum critical point is replaced with a quantum tricritical point and $\psi^{tri}$ determines the shape of the whole transition line in the vicinity of $T=0$ (for $|a_2-a_2^{(0)}|\ll \Lambda_0$).

\section{Summary}
In this work we applied the $\phi^6$ model to analyze the effect of thermal and quantum fluctuations on the phase diagram of a system of itinerant fermions exhibiting a quantum phase transition at wavevector $Q=0$. The analysis is restricted to discrete symmetry breaking and relies on an effective bosonic action analogous to that proposed by Hertz. \cite{hertz76} Renormalization of the action is computed from a system of flow equations derived by truncating the exact functional RG flow equation in the one particle irreducible scheme. The flow equations capture quantum and thermal fluctuations on equal footing and are applicable for $T\geq 0$ in $d=2,3$, also in the vicinity of the first or second order transition line. 

By analyzing the renormalized phase diagram at $T=0$ we find that a quantum critical point may be realized even if the bare action corresponds to a first order transition. Such scenario occurs for slightly negative quartic coupling $a_4$. 

We analyzed the dependence of the tricritical quartic coupling $a_4^{tri}$ as a function of temperature. The obtained decreasing form of $a_4^{tri}(T)$ implies enhancing the tendency towards a second order transition upon increasing the temperature. Therefore, if the transition is first order at $T=0$, it becomes second order at higher temperatures, namely for $T>T^{tri}$. 

We computed phase diagrams in the variables $(a_2,T)$, where $a_2$ acts as a non-thermal control parameter. In $d=2$ and $d=3$ we considered cases where the quantum phase transition is first ($a_4<a_4^{tri}$) and second ($a_4>a_4^{tri}$) order. In the former situation, the phase boundaries in $d=2,3$ show linear behavior in the vicinity of $T=0$, where the slope in $d=2$ is infinite (or very large). Approximately linear behavior of the $T_c$-line is also recovered for $a_4>a_4^{tri}$ in $d=2$.

Subsequently, we analyzed the case $a_4>a_4^{tri}$ in $d=3$. Numerical computations of the $T_c$ line for different choices of $a_4$ reveal that the phase boundary follows a power law with the shift exponent $\psi=3/4$, as predicted by Millis, \cite{millis93} as long as $a_4$ is separated from $a_4^{tri}(T=0)$. Upon decreasing $a_4$ towards $a_4^{tri}(T=0)$, the region in the phase diagram, where this scaling is observed shrinks. This is accompanied by formation of a different scaling region (at slightly higher $T$), where $\psi=\psi^{tri}=3/8$. When $a_4=a_4^{tri}(T=0)$ the quantum critical point is replaced by a quantum tricritical point and the shape of the whole transition line in the vicinity of $T=0$ is described by $\psi^{tri}$. These observations are understood by invoking scaling analysis relying on mean-field like assumption, by which the Ginzburg and transition lines in the vicinity of the quantum critical point are identified with each other. By deriving a formula for $\psi$ as function of the dynamical exponent $z$ and scaling dimension $[a_n]$ of the most relevant coupling, we recover $\psi=3/4$ for quantum criticality in $d=z=3$, and $\psi=3/8$ for quantum tricriticality.  

An interesting avenue for future research would be to reconsider these results in the context of specific microscopic fermionic models, from which coefficients of the Landau expansion can possibly be extracted as functions of physical parameters like $T$ and the chemical potential $\mu$ for instance. One could then check, in what conditions quantum criticality (and tricriticality) induced by order parameter fluctuations is conceivable, i.e. one encounters $a_4^{tri}\leq a_4<0$  in the vicinity of the transition. An interesting candidate for such a study is the so called $f$-model \cite{metzner03} of discrete symmetry-breaking Fermi surface deformations, for which $z=3$, and which for small $T$ and within mean field approximation, exhibits a phase diagram qualitatively similar to the one obtained here in Fig.~\ref{phase_diag_d2_Temp} (b). \cite{yamase05} 

\begin{acknowledgments}
The author would like to thank  H.W. Diehl, A.A. Katanin, A. Maciolek, W. Metzner, M. Napi\'{o}rkowski, P. Strack, and H. Yamase for very useful discussions, and J. Bauer and S. Takei for providing valuable comments on the manuscript. The support from the German Science Foundation through the research group FOR 723 is gratefully acknowledged. 
\end{acknowledgments}

\end{document}